\documentclass{jfm}
\usepackage{graphicx}
\usepackage{epstopdf, epsfig}

\usepackage{hyperref}
\usepackage{amsmath}
\usepackage{amssymb}
\usepackage{subfigure}

\title{Generalised and efficient wall boundary condition treatment in GPU-accelerated smoothed particle hydrodynamics}

\author{Massoud Rezavand\aff{1},
	Chi Zhang\aff{1}
	\and Xiangyu Hu\aff{1}
	\corresp{\email{xiangyu.hu@tum.de}}}

\affiliation{\aff{1}Department of Engineering Physics and Computation, 
	Technical University of Munich, Boltzmannstr. 15, 85748 Garching, Germany
}

\begin{document}
	\maketitle
	
	\begin{abstract}
		This paper presents a generalised and efficient wall boundary treatment in the smoothed particle hydrodynamics (SPH) method 
		for 3-D complex and arbitrary geometries with single- and multi-phase flows to be executed on graphics processing units (GPUs).
		Using a force balance between the wall and fluid particles with a novel penalty method, 
		a pressure boundary condition is applied on the wall dummy particles which 
		effectively prevents non-physical particle penetration into the wall boundaries 
		also in highly violent impacts and multi-phase flows with high density ratios. 
		A new density reinitialisation  
		scheme is also presented to enhance the accuracy.
		The proposed method is very simple in comparison with previous wall boundary formulations on GPUs
		that enforces no additional memory caching and thus is
		ideally suited for heterogeneous architectures of GPUs.
		The method is validated in various test cases involving violent single- and multi-phase 
		flows in arbitrary geometries and demonstrates very good robustness, accuracy and performance. 
		The new wall boundary condition treatment is able to improve the high accuracy of its previous 
		version \citep{ADAMI2012wall} also in complex 3-D and multi-phase problems, while it is efficiently executable
		on GPUs with single precision floating points arithmetic which makes it suitable for a wide range of 
		GPUs, including consumer graphic cards. Therefore, the method is a reliable solution for the 
		long-lasting challenge of the wall boundary condition in the SPH method
		for a broad range of natural and industrial applications.
		
	\end{abstract}
	

\section{Introduction}
Due to the peculiarities of the smoothed particle hydrodynamics (SPH) method to cope with
free-surface flows, complex geometries, multi-phase flows and many other 
natural or industrial problems, this method has attracted a considerable deal of attention
in recent years \citep[see e.g.][]{luo2021particle,KHAYYER2021103342,shimizu2020enhanced, zhang2020integrative,hopp2019viscous,vazquez2019shear}. 
However, the contribution of SPH in solving industrial problems is
not overwhelming yet, as a large amount of work still has to be done
to address the unknown properties of SPH \citep{rezavand2019fully, Rahmat_dewatering2020}. As a well-known instance, 
the accurate treatment of the boundary conditions has always been a challenge \citep{ADAMI2012wall,valizadeh2015study} 
and thus is highlighted as one of the SPH grand challenges \citep{vacondio2020grand}. As the applicability of SPH solvers 
 in industrial problems highly depends on their computational efficiency, the treatment of boundary conditions on 
parallel architectures, in particular graphics processing units (GPUs), becomes increasingly more important.

Apart from the reflective boundary condition \citep{fraga2019implementation} that has been recently proposed, 
the widely applied wall boundary conditions in the SPH community can be divided into two general groups. 
In the first group \citep[see e.g.][]{libersky1993high, monaco2011sph, ADAMI2012wall}, 
a set of fictitious particles are used for 
modelling the wall boundaries to ensure that the support domain of the smoothing kernel function is 
covered by enough number of particles. The enhanced variants of the methods belonging 
to this group are suitable to realise complex 2-D and 3-D geometries \citep{ADAMI2012wall}, as well as multi-phase flows 
\citep{rezavand2020weakly,rezavand2018isph}, while being ideally suited to parallel architectures like GPUs
\citep[see e.g.][]{WINKLER2017165, peng2019loquat}. In the second group, either surface integrals are 
introduced when the physical quantities are computed close to the boundaries 
\citep[see e.g.][]{kulasegaram2004variational, marongiu2010free, mayrhofer2015unified}, or artificial 
repulsive forces like a Lennard-Jones potential force are exerted on the fluid particles 
to prevent them crossing the boundaries \citep[see e.g.][]{monaghan1994simulating, monaghan2009sph}. 
Although using the surface integrals leads to consistent solutions
and acceptable results, the extension to 3-D geometries or multi-phase flows is not straightforward and the normal 
vectors of the boundaries have to be computed at each time step for flexible boundaries 
\citep{valizadeh2015study}. On the other hand, 3-D complex geometries can be easily
generated using the artificial repulsive forces while it may violate the kernel truncation 
in the immediate vicinity of the wall boundaries. These are due to the fact that in the models 
of the second group only a single layer of particles is required to mimic the boundaries, which
is an advantageous characteristic.

The theoretical aspects of solid wall models in the weakly compressible SPH (WCSPH) have been comprehensively studied by \citet{valizadeh2015study},
where several widely applied models are considered. They concluded that the method proposed 
by \citet{ADAMI2012wall} together with the density diffusion terms of \citet{antuono2012numerical} obtains
the most satisfactory agreement with experiments 
among the discussed formulations in their work.
As the SPH computations are inherently expensive, the efficiency of boundary treatment,
as a crucial aspect of numerical analysis, on heterogeneous architectures is of substantial importance.  
In the context of GPU-accelerated SPH solvers, all the aforementioned wall boundary conditions are 
implemented and analysed  \citep[see e.g.][]{herault_sph_2010, crespo2015dualsphysics, cercos2015aquagpusph, 
WINKLER2017165, alimirzazadeh2018gpu, peng2019loquat}. However, the conclusions of \citet{valizadeh2015study}
are also reflected in this context and the majority of these solvers benefit from the methods of the 
first group, in particular the model proposed by \citet{ADAMI2012wall}, which shows that 
this method is the ideal choice for GPU-accelerated SPH solvers. 

The above reflections are the motivations of the 
present study to generalise the work of \citet{ADAMI2012wall} to 3-D complex geometries and 
multi-phase problems, which are not yet extensively addressed according to the existing literature
\citep{valizadeh2015study,rezavand2020weakly}. In this work we address this long-lasting open challenge of 
the SPH method and propose a novel wall boundary condition which ideally deals with complex 3-D geometries 
and multi-phase problems with abrupt physical discontinuities at the multi-phase interface and when
interacting with wall boundaries. The method is also implemented on the heterogeneous architecture of
GPUs to propose a reliable solution for a wide range of natural and industrial problems.

The proposed wall treatment achieves highly satisfactory results without a need for particle regularisation methods
or accurate calculation of the wall normal vectors. 
On the other hand, to obtain accurate results the new methodology does not need density diffusion terms, which are not able to be extended in a straightforward way for 
multi-phase flows and are deemed to be necessary by \citet{valizadeh2015study} to 
obtain the most satisfactory results with the method of 
\citet{ADAMI2012wall}. We therefore believe that the present method can be a reliable solution for 
the long-lasting challenge of the wall boundary condition in the SPH method
for a broad range of natural and industrial applications. 
The present method uses the estimates of the outward wall unit
normal vectors with no need for its accurate calculation \citep{rezavand2020weakly,zhang2017weakly} and 
thus enforces no additional memory caching which is ideally suited to GPU architectures and therefore more efficient. 
Using a force balance between the wall and fluid particles with a novel penalty method, 
a pressure boundary condition is applied on the wall dummy particles that 
effectively prevents particle penetration also in highly violent impacts and 
multi-phase flows with high density ratios.
In addition, we introduce a new density reinitialisation technique
to enhance the accuracy and robustness. Various 2-D and 3-D test cases are considered to asses the 
accuracy and robustness of the proposed method and the consistency of the formulation 
is also analysed from several aspects. 

The rest of this paper is organized as follows: section \ref{sec:NumericalModel} details the theoretical
aspects of the proposed wall boundary conditions. The considerations for
the implementation of the proposed method on GPUs are next explained in section \ref{sec:gpuImplementation}.
The numerical results are then presented 
and discussed in section \ref{sec:ResultsandDiscussion} and finally, the 
concluding remarks of the present study are summarised in section \ref{sec:conclusions}. 
All the computational codes and data-sets accompanying this work are released 
in the repository of SPHinXsys \cite{zhang2020sphinxsys, zhang2021sphinxsys} at \url{https://www.sphinxsys.org}.

\section{Numerical method}\label{sec:NumericalModel}
\subsection{Governing equations}
For an inviscid flow
the mass and momentum conservation equations can be 
written respectively as
\begin{equation}\label{eq:massGeneral}
	\frac{\mathrm{d}\rho}{\mathrm{d}t} = -\rho\nabla\cdot\mathbf{v},
\end{equation}
and
\begin{equation} \label{eq:momentumGeneral}
	\frac{\text{d}\mathbf{v}}{\text{d}t} = -\frac{1}{\rho}\nabla p + \mathbf{g} + \mathbf{F}_p,
\end{equation}
where $\frac{\mathrm{d}}{\mathrm{d}t}$ is the material or Lagrangian derivative, $\rho$ the density, 
$\mathbf{v}$ the velocity, $p$ the pressure, $\mathbf{g}$ 
the gravitational acceleration and $\mathbf{F}_p$ a new force balance term obtained from a penalty method. 
Note that $\mathbf F_p$ is applied only when a wall particle is interacting 
with a light phase (air) particle in multi-phase flow with high density ratios to 
prevent particle penetration into the wall boundaries. 
To close the system, pressure is estimated from density via an artificial equation of state, 
within the weakly compressible regime. Here, 
we use a simple linear equation for both the heavy and light phases
\begin{equation}\label{eq:EoS}
	p = c^{2}(\rho-\rho^0),
\end{equation} 
where $c$ and $\rho^0$ are the speed of sound and the initial reference density. 
In this study, we assume that the speed of sound is constant 
and we set $c=10U_{max}$, where $U_{max}$ denotes the 
maximum anticipated velocity of the flow. 

Within the WCSPH framework, there are two different formulations to implement the 
conservation of mass, 
viz. the particle summation and continuity equation \citep{Monaghan_2012_Annual_Rew}. 
The former calculates density through a summation over all the neighbouring particles
\begin{equation}\label{eq:densitySummation}
	\rho_i = m_i\sum_{j}W_{ij},
\end{equation}
where $m_i$ denotes the mass of particle $i$ and the smoothing kernel function 
$ W(\left| \mathbf{r}_{ij}\right| ,h)$ is simply substituted by $ W_{ij} $, 
with $\mathbf{r}_{ij}=\mathbf{r}_{i}-\mathbf{r}_{j}$ being the displacement vector between particle $i$ and $j$ 
and $h$ the smoothing length, 
respectively.
The latter updates particle density by discretising the continuity equation as
\begin{equation} \label{eq:SPH_massConservation}
	\frac{\text{d}\rho_{i}}{\text{d}t} = \rho_i \sum_{j}\frac{m_j}{\rho_j}\mathbf{v}_{ij}\cdot\nabla_{i} 
	W_{ij}=2\rho_i\sum_{j} \frac{m_j}{\rho_j} (\mathbf{v}_{i}- \mathbf{\overline{v}}_{ij}) 
	\cdot\nabla_{i} W_{ij},
\end{equation}
where $\mathbf{v}_{ij}=\mathbf{v}_{i}-\mathbf{v}_{j}$ is the relative velocity 
and $\mathbf{\overline{v}}_{ij}=\frac{\mathbf{v}_{i}+\mathbf{v}_{j}}{2}$ denotes
the average velocity between particle $i$ and $j$. In the two-phase cases of 
the present work, we follow \citet{rezavand2020weakly} and use Eqs. \eqref{eq:densitySummation} and \eqref{eq:SPH_massConservation} for the lighter phase
and the denser phase, respectively. For the single-phase cases we use Eq.\eqref{eq:SPH_massConservation} for the entire domain including the boundary particles.

Following \citet{Monaghan_2012_Annual_Rew},
excluding the artificial viscosity, 
the momentum conservation equation can be discretised as
\begin{equation} \label{eq:SPH_momentumConservation}
	\frac{\text{d}\mathbf{v}_{i}}{\text{d}t} = -\sum_{j}m_j \left( \frac{p_i + p_j}{\rho_i\rho_j}\right)
	\nabla_{i} W_{ij}=-2\sum_{j} m_j \frac{\overline{{p}}_{ij}}{\rho_i\rho_j} \nabla_{i} W_{ij},
\end{equation}  
with $\overline{{p}}_{ij}=\frac{p_{i}+p_{j}}{2}$ being the average pressure between 
particle $i$ and $j$. 

\subsection{WCSPH based on a Riemann solver}
\label{subsec:RiemannbasedSPH}
In the SPH methods based on the Riemann solvers \citep{vila1999particle,moussa2006convergence, zhang2017weakly}, 
an inter-particle Riemann problem is constructed 
along the unit vector $\mathbf{e}_{ij}=-\frac{\mathbf{r}_{ij}}{\left| \mathbf{r}_{ij}\right|}$, 
with the initial left and right states reconstructed 
from particle $i$ and $j$ by piecewise constant approximation, respectively, as
\begin{equation}
	\left. \begin{array}{ll}
	
			\displaystyle (\rho_L, U_L, p_L) = (\rho_i, \mathbf{v}_i \cdot \mathbf{e}_{ij}, p_i)\\[8pt]
	
			\displaystyle(\rho_R, U_R, p_R) = (\rho_j, \mathbf{v}_j \cdot \mathbf{e}_{ij}, p_j)
	
		\end{array}\right\},
	\label{eq:LRstates}
\end{equation}
where subscripts $L$ and $R$ denote left and right states, respectively, and the 
discontinuity is assumed at $\mathbf{\overline{r}}_{ij}=\frac{\mathbf{r}_{i}+\mathbf{r}_{j}}{2}$.
For more details about the construction and solution of the single- and multi-phase Riemann problem, 
the readers are referred to \citet{zhang2017weakly} and \citet{rezavand2020weakly}, respectively.

We approximate the intermediate velocity and pressure as
\citep{toro1989fast,zhang2017weakly, rezavand2020weakly}
\begin{equation}
	\left. \begin{array}{ll}
	
			\displaystyle U^* = \overline{U} +\frac{1}{2}\frac{p_L - p_R}{c\overline{\rho}}\\[8pt]
	
			\displaystyle p^* = \overline{P} + \frac{\rho_L\rho_R\beta(U_L - U_R)}{(\rho_L+\rho_R)}
	
		\end{array}\right\},
	\label{eq:RSsolution}
\end{equation}
where $\overline{\rho}=\frac{\rho_L  + \rho_R }{2} $, $\overline{U} =\frac{\rho_L U_L + \rho_R U_R}{\rho_L+\rho_R}$ 
and $\overline{P} =\frac{\rho_L p_R + \rho_R p_L}{\rho_L+\rho_R}$.
In Eq. (\ref{eq:RSsolution}), 
the velocity and pressure solutions consist of a density-weighted average term and a
 numerical dissipation term.
Due to the assumption of weak compressibility, 
the magnitudes of the second terms (numerical dissipation) are much smaller than that of the first term. 
To reduce the numerical dissipation,
a limiter $\beta= \mathrm{min}(3 \mathrm{max} (U_L - U_R,0), c)$ 
is applied for $p^*$ in Eq. \eqref{eq:RSsolution} \citep{zhang2017weakly,rezavand2020weakly}.
This results in a low-dissipation numerical scheme
which will be further analysed in section \ref{subsec:twophase_staticTank}.

Having the intermediate values determined, the mass and  momentum conservation equations, 
i.e. Eqs. \eqref{eq:SPH_massConservation} and \eqref{eq:SPH_momentumConservation}, 
can be rewritten as
\begin{equation} \label{eq:SPH_star_massConservation}
	\frac{\text{d}\rho_{i}}{\text{d}t} = 2\rho_i\sum_{j} \frac{m_j}{\rho_j} (\mathbf{v}_{i}- \mathbf{v}^{*}) \cdot\nabla_{i} W_{ij},
\end{equation}
and
\begin{equation} \label{eq:SPH_star_momentumConservation}
	\frac{\text{d}\mathbf{v}_{i}}{\text{d}t} = -2 \sum_{j} m_j \frac{{p}^{*}}{\rho_i \rho_j} \nabla_{i} W_{ij},
\end{equation}
where $\mathbf{v}^{*} = U^{*} \mathbf{e}_{ij} + (\frac{\rho_i \mathbf{v}_i + \rho_j \mathbf{v}_j}{\rho_i+\rho_j} - \overline{U}\mathbf{e}_{ij} )$. 
The term $\frac{\rho_i \mathbf{v}_i + \rho_j \mathbf{v}_j}{\rho_i+\rho_j}$ was introduced
by \citet{rezavand2020weakly}
for the sake of consistency and enhances the
formulations in comparison to the work of
\citet{zhang2017weakly}, which used $\mathbf {\overline{v}}_{ij}$ instead. 
It is worth noting that Eq. \eqref{eq:SPH_star_momentumConservation} is also an antisymmetric 
relation and conservative.

\subsection{Efficient wall boundary condition}
\label{subsec:wallBoundary}
As mentioned before, \citet{valizadeh2015study}
concluded in a comprehensive study that the wall boundary model proposed 
by \citet{ADAMI2012wall} together with a density diffusion term
is in the very satisfactory agreement with experiments and 
is the best method among the discussed formulations in their work.
Here we further improve the work of \citet{ADAMI2012wall}
for highly dynamic complex 3-D and multi-phase flows characterised by high density ratios
to be executed on GPUs. The present formulation does not need the 
density diffusion terms deemed necessary by \citet{valizadeh2015study}. 

\begin{figure}
	\centering
	\includegraphics[width=0.9\textwidth]{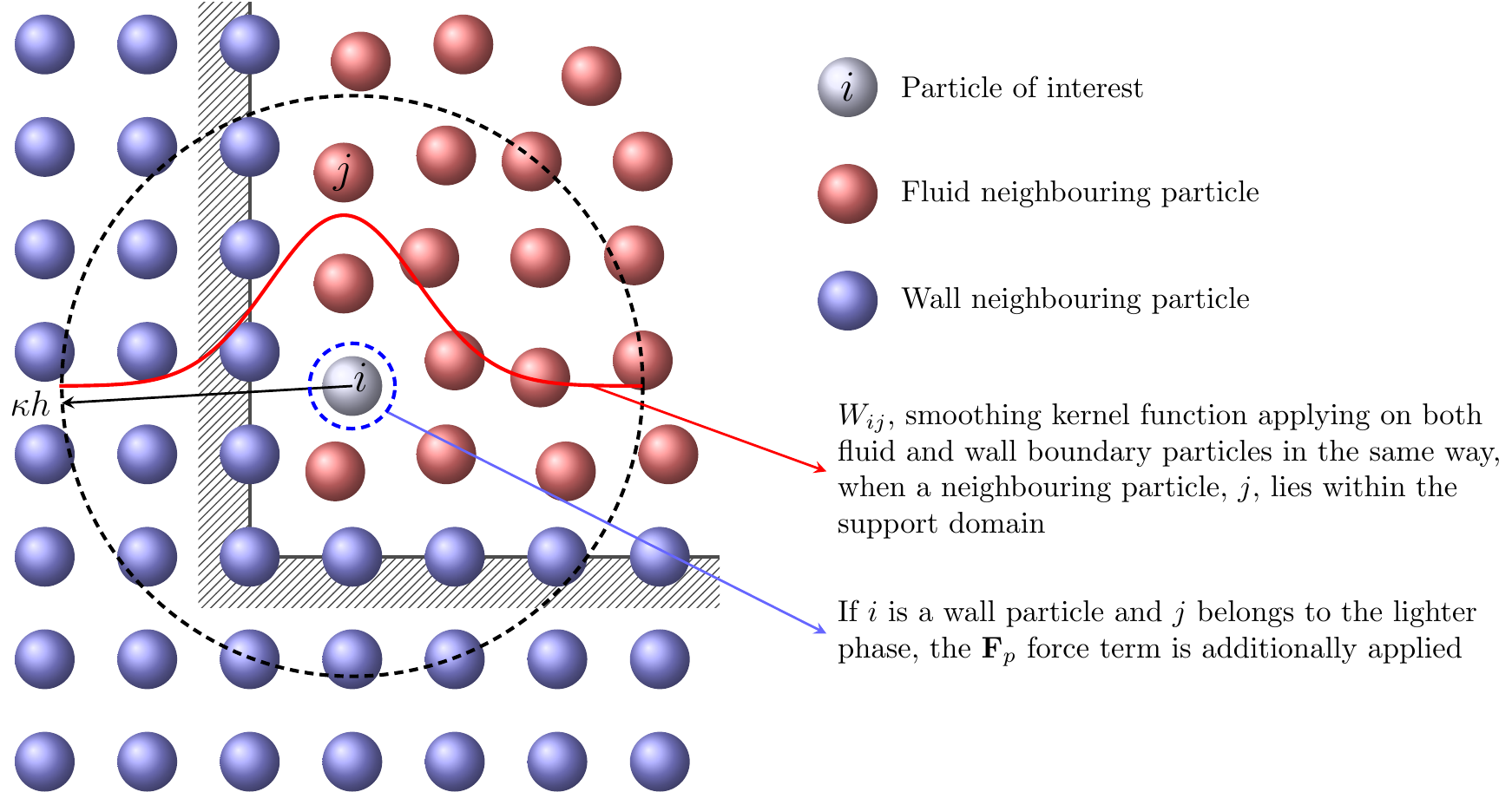}
	\caption{Schematic illustration of the wall boundary condition: The blue particles represent
		the dummy wall boundaries and are fixed during the entire simulation. Fluid particles (in red) 
		interact with the dummy particles and neighbouring fluid particles that lie in the support 
		domain. In this work we use a $5th$-order Wendland kernel \citep{Wendland1995}
		for which $\kappa = 2$. }
	\label{fig:BCschematic}
\end{figure}

Here, we use dummy particles to impose the solid wall 
condition as shown in figure \ref{fig:BCschematic}. 
The main advantages of dummy particles
compared to other wall boundary treatments mentioned 
earlier (e.g. the methods using surface integrals) 
is the simplicity when dealing with complex geometries, and also that the boundary is well-described throughout
the simulation once the particles have been generated, which are of importance in the context of 
GPU-accelerated and high performance SPH solvers. 
From a practical point of view, this characteristic makes it possible to generate a separate data structure for the wall boundary particles on GPUs, which does not 
alter and can be reused during the whole simulation.
Such methodologies have already demonstrated considerable 
computational performance enhancements \citep[see e.g.][]{zhang2020dual}  

In figure \ref{fig:BCschematic}, the fluid particles (in red) near the wall interact with 
the wall dummy particles (in blue), which lie within the support radius of the smoothing kernel function, $W_{ij}$. 
As a consequence, the dummy wall particles contribute to the mass and momentum 
conservation equations the same as the fluid particles. 
In order to realise the fluid-wall interactions, we only need to solve the 
constructed Riemann problem with the following intermediate pressure between particles 
of fluids and wall dummy particles, 
the same as for fluid-fluid interactions (see section \ref{subsec:RiemannbasedSPH})

\begin{equation}\label{eq:RSatWall}
	p^* = \frac{\rho_f p_w + \rho_w p_f}{\rho_f+\rho_w},
\end{equation}
where subscripts $f$ and $w$ denote fluid and wall, respectively. 
This will decrease the wall-induced numerical dissipation. 
It is worth mentioning that the intermediate velocity value $U^*$ is still obtained through Eq. \eqref{eq:RSsolution}, 
as for fluid-fluid interactions, depending on the problem being single- or multi-phase. 
We next calculate the pressure of the
wall dummy particles by a summation over all contributions of the 
neighbouring fluid particles, similar to \citet{ADAMI2012wall}
\begin{equation}\label{eq:wallPressure}
	p_w = \frac{\sum_{f} p_f W_{wf} + (\mathbf{g}-\mathbf{a}_{w}) \cdot \sum_{f} \mathbf{r}_{wf} 
		W_{wf}}{\sum_{f} W_{wf}} ,
\end{equation}
where $\mathbf{a}_{w}$ is the wall acceleration.
Note that, for the multi-phase problems with high density ratios,
the above equation is divided by $\rho_f$ to vanish
the contribution of the heavy phase particles in $p_w$ at a triple point, 
where water, air and solid particles meet \citep{rezavand2020weakly}.
The density of wall dummy particles is then obtained from pressure using the
equation of state presented by Eq. (\ref{eq:EoS}).
Another important aspect of this formulation is that the
implementation of Eq. \eqref{eq:wallPressure} retains
the parallel nature of the algorithm related to the SPH method on GPUs 
\citep{WINKLER2017165,peng2019loquat}.
 
 \subsection{Penalty method}
 As mentioned before and seen in Eq. \eqref{eq:momentumGeneral}, 
 in the present methodology we introduce a novel force balance term obtained from a penalty method, 
 which effectively prevents particle penetration also in highly violent impacts and 
 multi-phase flows with high density ratios. As also
 explained in figure \ref{fig:BCschematic}, this force is only applied when the particles of 
 the lighter phase (e.g. $i\in$air) interact with wall particles ($j=w$) and reads 
 \begin{equation} \label{eq:SPH_penaltyMethodTerm}
 	\mathbf{F}_p = -2\sum_{w} \frac{m_w}{\rho_i\rho_w} \mathcal{P} \mathbf{n}_{w} \frac{1}{\left|\mathbf{r}_{iw}\right|^2} 
 	\frac{\partial W_{iw}}{\partial (\left|\mathbf{r}_{iw}\right|)},
 \end{equation}
where $\mathbf{n}_{w}$ is the estimate for the unit normal vector of the wall particles, 
$\frac{\partial W_{iw}}{\partial (\left|\mathbf{r}_{iw}\right|)}$ is the first derivative of the 
kernel function and 
$\mathcal{P} =\gamma_p \lambda |p_i \mathbf{e}_{iw} \cdot \mathbf{n}_{w}|$ is the penalty parameter for which
$\gamma_p$ denotes the penalty strength and
\begin{equation}
	\left. \begin{array}{ll}
		\displaystyle \lambda = \frac{(1-\delta)^2}{dx}
		\quad \mbox{on\ }\quad \delta< 1 \\[8pt]
		\displaystyle  \lambda = 0
		\quad \mbox{on\ }\quad \delta\geq 1
	\end{array} \right\},
	\label{symbc}
\end{equation}
where $dx$ is the initial particle spacing and
\begin{equation} \label{eq:deltaForPenalty}
	\delta = \frac{2\left|\mathbf{r}_{iw}\right|}{dx} (\mathbf{e}_{iw} \cdot \mathbf{n}_{w}).
\end{equation}
In the simulations of the present study $\gamma_p=1$ and the unit normal vector for wall 
particles are calculated by \citep{randles1996smoothed, zhang2017weakly}
\begin{equation} \label{eq:normalVectors}
	\mathbf{n}_{w} = \frac{\Psi(\mathbf{r}_i)}{\left|\Psi(\mathbf{r}_i)\right|}, 
	\quad \mbox{with\ }\quad \Psi(\mathbf{r}_i) = -\sum_{j\in w} \frac{m_j}{\rho_j}\nabla_{i} W_{ij}, 
\end{equation}
where the summation is only applied on the wall particles. 
 
 \subsection{Density reinitialisation}
 In order to further enhance the robustness and accuracy of the proposed method, 
 we propose a simple and efficient formulation to reinitialise the density field. 
 In this formulation, the 
 density field of fluid particles in free-surface flows is reinitialised at every time step using the following formulation
 \begin{equation} \label{eqrhosum}
 	\rho_i = \rho^0 \frac{ \sigma_i }{\sigma^0_i } +  \max(0, ~\rho^* - \rho^0 \frac{ \sigma_i }{\sigma^0_i }) \frac{\rho^0}{\rho^*},
 \end{equation}
 where $\sigma_i = \sum W_{ij}$ and $\sigma^0_i = \sum W^0_{ij}$ 
 are the current and the initial number particle densities, respectively, 
 $\rho^*$ denotes the density before reinitialisation and superscript $0$ represents the initial reference value.
 For flows without free surface, Eq.  (\ref{eqrhosum}) is simplified as 
 \begin{equation} \label{eqrhosumnosurface}
 	\rho_i =  \rho^0 \frac{ \sigma_i}{\sigma^0_i }, 
 \end{equation}
 for internal flow. 
 Note that the light phase experiences the heavy phase like moving wall boundary, 
 while the heavy phase undergoes a free-surface-like flow with variable free-surface pressure 
 in present multi-phase model. 
 Therefore, 
 Eqs. \eqref{eqrhosum} and \eqref{eqrhosumnosurface} are applied to light and heavy phases, 
 respectively. 
 \subsection{Time integration}
 At the beginning of every time step, the fluid density field is initialised 
 by Eq.\eqref{eqrhosum} or Eq.\eqref{eqrhosumnosurface}.
 To integrate the equations of motion in time, the kick-drift-kick 
 \citep{monaghan2005,adami2013transport} scheme is employed. 
 The first half-step velocity is obtained as
 \begin{equation}
 	\mathbf{v}_i^{n + \frac{1}{2}} = \mathbf{v}_i^n +  \frac{\delta t}{2}  \big( \frac{\text{d} \mathbf{v}_i}{\text{d}t} \big)^{n},
 \end{equation}
 from which we obtain the position of particles at the next time step
 \begin{equation}
 	\mathbf{r}_i^{n + 1} = \mathbf{r}_i^n +  \delta t  \mathbf{v}_i^{n + \frac{1}{2}}.
 \end{equation}
 Having the reinitialised density field and new
 positions, we update the density field either with the
 continuity equation in Eq.\eqref{eq:SPH_massConservation} and the 
 following formulation for the heavy phases (e.g. water)
 \begin{equation} \label{eq:timeIntg_massConservation}
 	\rho_i^{n+1} = \rho_i^n + \delta t \big(\frac{\mathrm{d}\rho_{i}}{\mathrm{d}t}\big)^{n+ \frac{1}{2}},
 \end{equation}
 or solely with the summation formulation in \ref{eq:densitySummation} for the light phase (e.g. air).
 The pressure field is then updated using Eq.\eqref{eq:EoS}, accordingly.
 Now, we can obtain the time increment of the velocity field 
 by Eq.\eqref{eq:SPH_star_momentumConservation} and then update 
 the velocity of particles to the next time step as  
 \begin{equation}
 	\mathbf{v}_i^{n + 1} = \mathbf{v}_i^{n + \frac{1}{2}} +  \frac{\delta t}{2}  
 	\big( \frac{\text{d} \mathbf{v}_i}{\text{d}t} \big)^{n+1}.
 \end{equation}
 For numerical stability, the time step size is limited by the CFL condition
 \begin{equation}
 	\delta t \leq 0.25 \big( \frac{h}{c + U_{max}}  \big),
 \end{equation}
 and the body force condition
 \begin{equation}
 	\delta t \leq 0.25 \sqrt \frac{h}{\left|\mathbf{g} \right| }.
 \end{equation}
It is worth noting that following our previous work \citep{rezavand2020weakly},
also the present method uses the same speed of sound both for light and heavy phases which
considerably enhances the computational efficiency of the method. 
\section{GPU implementation}\label{sec:gpuImplementation}
In recent years the utilisation of the high floating point arithmetic performance
of GPUs has become increasingly more widespread for high performance scientific 
computing. 
With the heterogeneous multi-thread architectures of GPUs,
this new paradigm has made it possible to execute demanding large scale simulations
on general purpose computing workstations. 
One of the leading technologies in this scope
is the Nvidia graphic cards that are able to harness the computational power of 
thousands of GPU cores using the compute unified device architecture (CUDA).

In this study, we have implemented the proposed methodologies in CUDA using GPU well-suited data structures. 
Our numerical algorithm consists of the following main compartments: 
a) mass and momentum conservation equations, b) density update, c) pressure calculation for 
the dummy boundary particles, d) neighbour search and e) time 
integration. All of these multiple parts are implemented as CUDA kernels to be execute fully
on the device (GPU), thereby minimizing the costly communications between host (CPU) and 
device memories.
We only need to copy the simulation data once at the beginning from host to device, and 
vice versa when we need to save the computational results onto host at desired time intervals.
The method is fully developed on single precision floating points arithmetic, which makes it
possible to be efficiently executable even on consumer graphic cards.

Table \ref{tab:compsPercentage} shows the percentage of the computational effort for all the aforementioned compartments according to the profiling results
obtained by the \verb|Nvidia nvprof| profiler for a 3-D
multi-phase sloshing tank problem with $243~432$ particles. 
As expected, the most resource intensive parts are the mass 
and momentum equations. It is noted that among the corresponding functions of the 
proposed wall boundary condition treatment, only the pressure calculation for wall particles
and the density update formulation, respectively with 8.35\% and 5.94\%,
take a considerable percentage of computational time. The computational effort
corresponding to the penalty method and the normal vector calculation kernels are negligible.
This demonstrates the efficiency of the 
proposed wall model on the GPU architectures, as the core kernels (penalty method and normal vectors) 
introduce almost no additional computational overhead and the proposed method is thus evidently computationally
efficient and also well suited to GPU architectures as it also does not expose additional memory caching.

In our framework, we use a single data structure for both fluid particles and the wall dummy
particles. This is possible because, as we explained in section \ref{subsec:wallBoundary},
the dummy particles contribute in the mass and momentum conservation equations 
in the same way as fluid particles. For this reason there is no need to store additional 
data, enforced by the wall boundary model, on the limited shared memory per streaming multiprocessor (SM), 
which is necessary in previous wall boundary models for GPUs
\citep[see e.g.][]{fourtakas2019local}. 
The above-mentioned characteristics also imply that the number of 
operations per data element access on the device memory is quite large and it tends to be a 
compute-bound kernel rather a memory-bound/bandwidth-bound kernel \citep{lee2012cuda,alimirzazadeh2018gpu}. 
Having in mind that the computational power of the 
GPUs is already quite high and is still rapidly growing, i.e. the number of registers and size of caches are increasing, 
this will be of high importance also for further optimizations in the near future.
Similar advantages have been also observed in other GPU accelerated
SPH solvers employing the wall boundary model proposed by \citet{ADAMI2012wall} and running
on single GPUs \citep[see e.g.][]{WINKLER2017165,peng2019loquat}.
The proposed method in the present study retains these advantages, while enhancing the accuracy, robustness and
efficiency of the method of \citet{ADAMI2012wall}.
\begin{table}
	\centering
	\begin{tabular}{ cc}
		CUDA kernel   &  Time (\%) \\ 
		\hline
		Mass conservation (Riemann solver)			& 47.64 \\	
		Momentum conservation (Riemann solver)		& 34.96 \\
		BC pressure				& 8.35 \\
		Density update 		& 5.94 \\	
		Neighbour search (including sorting algorithms)		& 2.31 \\
		Time integration 	 & 0.64 \\		
		Other functions		& 0.16 \\
	\end{tabular}
\caption{Percentage of the computational effort for different CUDA kernels for a 3-D
	multi-phase sloshing tank problem.}
	\label{tab:compsPercentage}
\end{table}

\section{Numerical tests}\label{sec:ResultsandDiscussion}
In this section we first consider a two-phase hydrostatic tank 
with a high density ratio to assess the accuracy of the proposed method in static conditions.
Other test cases including a two-phase sloshing tank
and two single-phase dam break flows with complex arbitrary obstacles
are also considered to validate the proposed method 
for modelling highly violent flows with high density ratio and to analyse the robustness of the
proposed generalised method in complex 3-D problems, as well as challenging single-phase flows. 
We have used the quintic Wendland kernel \cite{Wendland1995}  
with a smoothing length of $h = 1.3 dx$ and a support radius of $2.6dx$. 
To set the artificial speed of sound $c$, the maximum velocity is 
estimated as $U_{max} = 2\sqrt{gH}$ according to the shallow-water theory 
\cite{ritter1892fortpflanzung}, where $H$ is the initial depth of the denser phase
and $g=9.81~\mathrm{m.s^{-2}}$ is the gravity acceleration for all the cases. 

For the execution of the simulations we have used a workstation with an 
Intel(R) Xeon(R) CPU E5-2603 v4 equipped by 128 GB RAM and a 
Nvidia GeForce RTX 2080 Ti graphics card.

\subsection{Two-phase hydrostatic tank}\label{subsec:twophase_staticTank}
\begin{figure}
	\centering
	\includegraphics[width=0.5\linewidth]{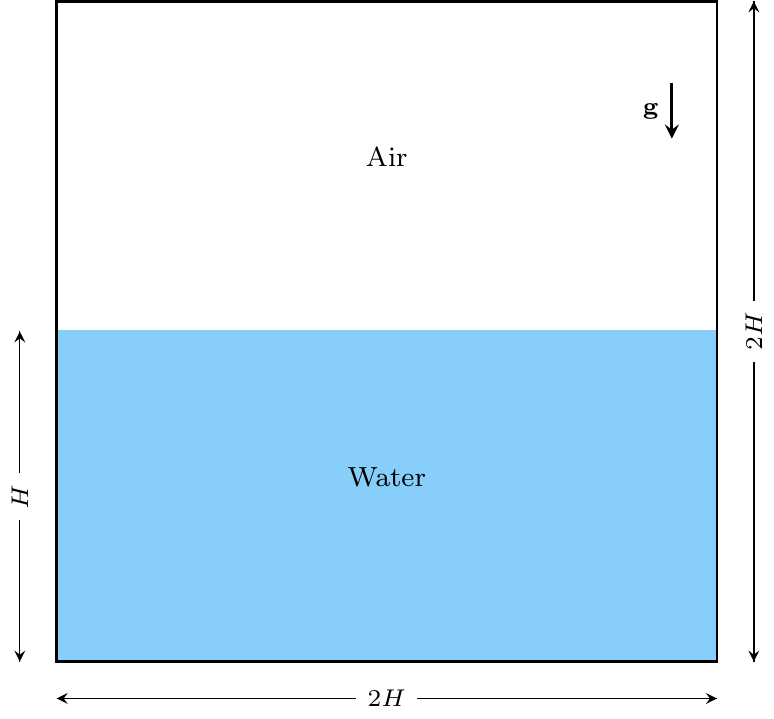}
	\caption{Two-phase hydrostatic tank: schematic illustration.}
	\label{fig:2phase_staticTank_config}
\end{figure}

\begin{figure}
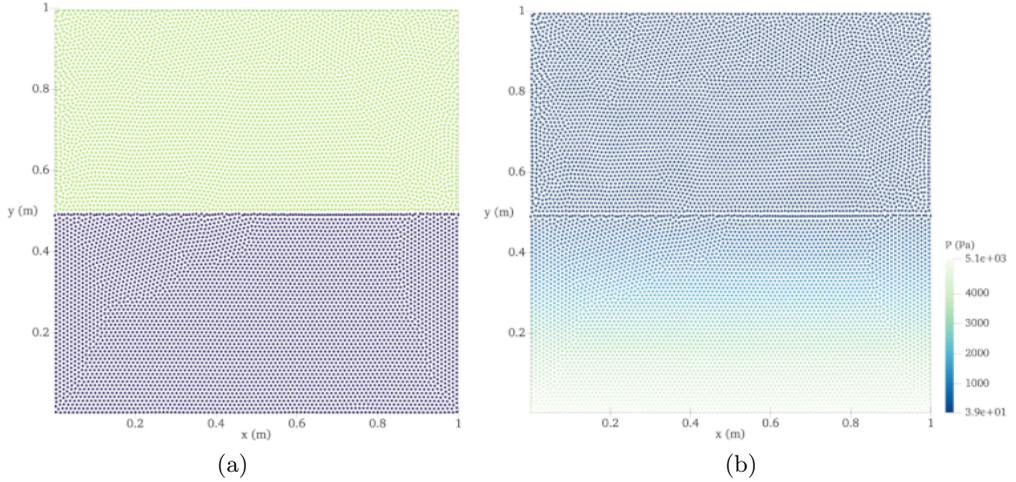

	\centering
	\subfigure []{\includegraphics[width=0.46\linewidth]{Figs/multiphase_hydrostaticTank/snapshots/twophaseInterface_t10_LQ-eps-converted-to.png}} 
	\subfigure []{\includegraphics[width=0.52\linewidth]{Figs/multiphase_hydrostaticTank/snapshots/pressureProfile_t10_LQ-eps-converted-to.png}}
	\caption{Two-phase hydrostatic tank: (a) particle distribution and 
		(b) pressure field at $t=10~$s.}
	\label{fig:staticTank_press_interface}
\end{figure}

A 2-D two-phase hydrostatic tank is first considered to evaluate the proposed method in static 
conditions in the presence of gravity. In this case we quantify energy conservation properties
of the proposed method, assess its accuracy 
and also analyse the convergence characteristics of the proposed boundary
condition. The 
lower half of the tank is filled with water and the upper half is 
occupied by air, both initially at rest. 
The schematic of this problem is illustrated in 
figure \ref{fig:2phase_staticTank_config}. 
The tank has a length of $L =2H=1~$m and the
initial water depth is $H = 0.5~$m. Initially, particles are
regularly placed on a Cartesian grid and to carry out convergence studies we use
four initial particle spacings, namely $dx = 0.01,~0.005,~0.0025$ and $0.00125~$m. 
The two fluids are considered to be inviscid and the density of water 
and air are $\rho_w = 1000~\mathrm{kg \cdot m^{-3}}$ and $\rho_a = 1~\mathrm{kg \cdot m^{-3}}$, respectively.
The obtained pressure contours and the particle distribution are depicted in figure \ref{fig:staticTank_press_interface} 
at $t=10~$s. 
It can be observed that a quite smooth pressure field and a stable two-phase interface is obtained 
without notable non-physical interfacial behaviour. It is also noteworthy that both the pressure 
field and particle distribution in the vicinity of the wall boundaries are also quite
uniform, which shows the effectiveness of the proposed wall boundary condition. 

\begin{figure}
	\centering
	\includegraphics[trim = 0mm 0mm 0mm 0mm, clip,width=0.45\linewidth]{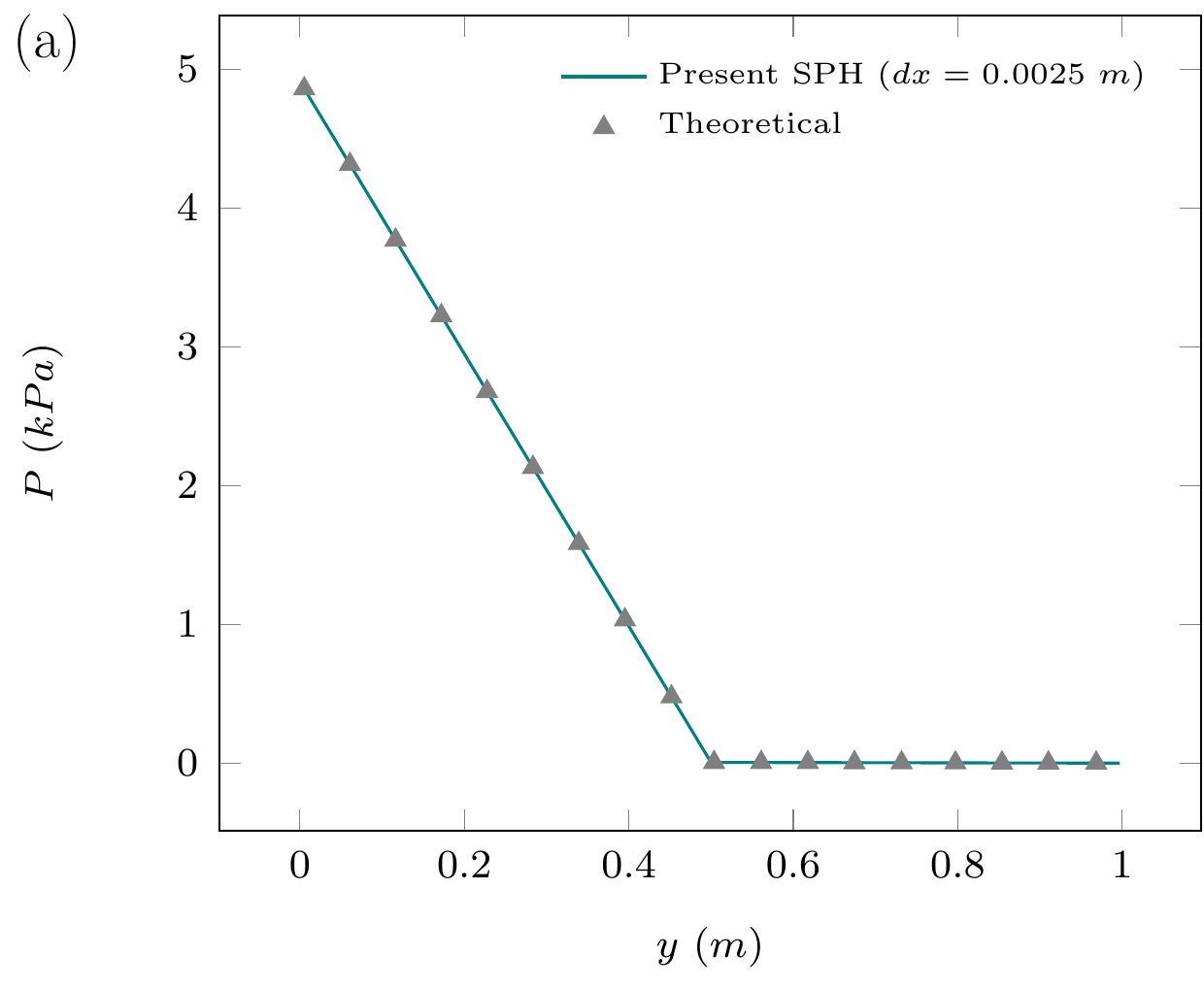}
	\hspace{0.3cm}
	\includegraphics[trim = 0mm 0mm 0mm 0mm, clip,width=0.45\linewidth]{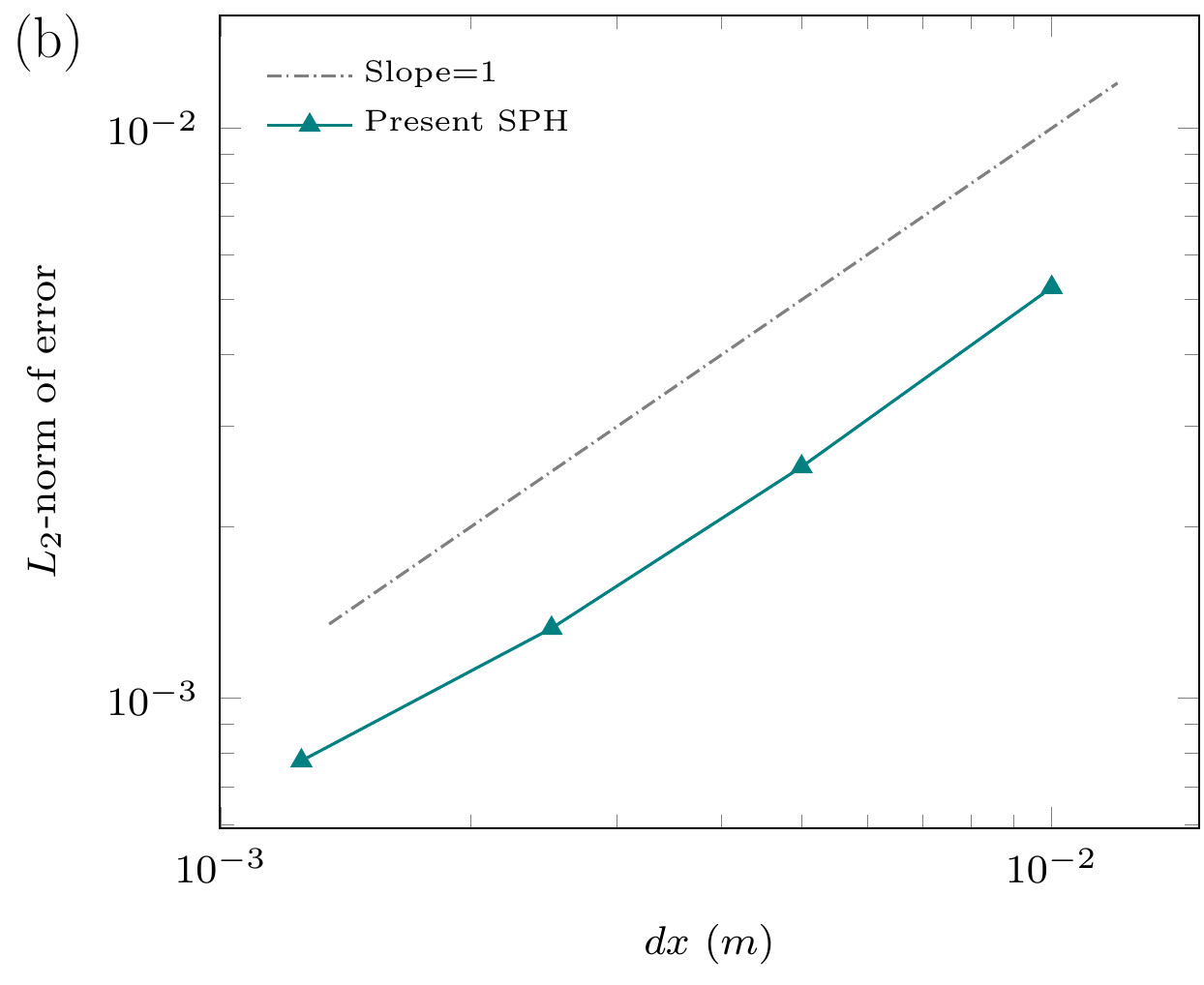}
	\caption{Two-phase hydrostatic tank: (a) comparison of the pressure profile with analytical solution and (b) $L_2$-norm convergence analysis at $t=10~$s.}
	\label{fig:staticTank_pressProfile_L2norm}
\end{figure}

The accuracy and convergence properties of the proposed methodology are next 
evaluated through the comparison of the numerically computed pressure profile
with the analytical hydrostatic pressure values. Figure \ref{fig:staticTank_pressProfile_L2norm} (a) compares 
the obtained pressure profile with the analytical solution
and figure \ref{fig:staticTank_pressProfile_L2norm} (b)
shows the convergence properties of the proposed method using the $L_2$-norm of error \citep{Fatehi_2011_error}. 
As it can be observed, 
an approximately first-order convergence is obtained.

\begin{figure}
	\centering
	\includegraphics[trim = 0mm 0mm 0mm 0mm, clip,width=0.45\linewidth]{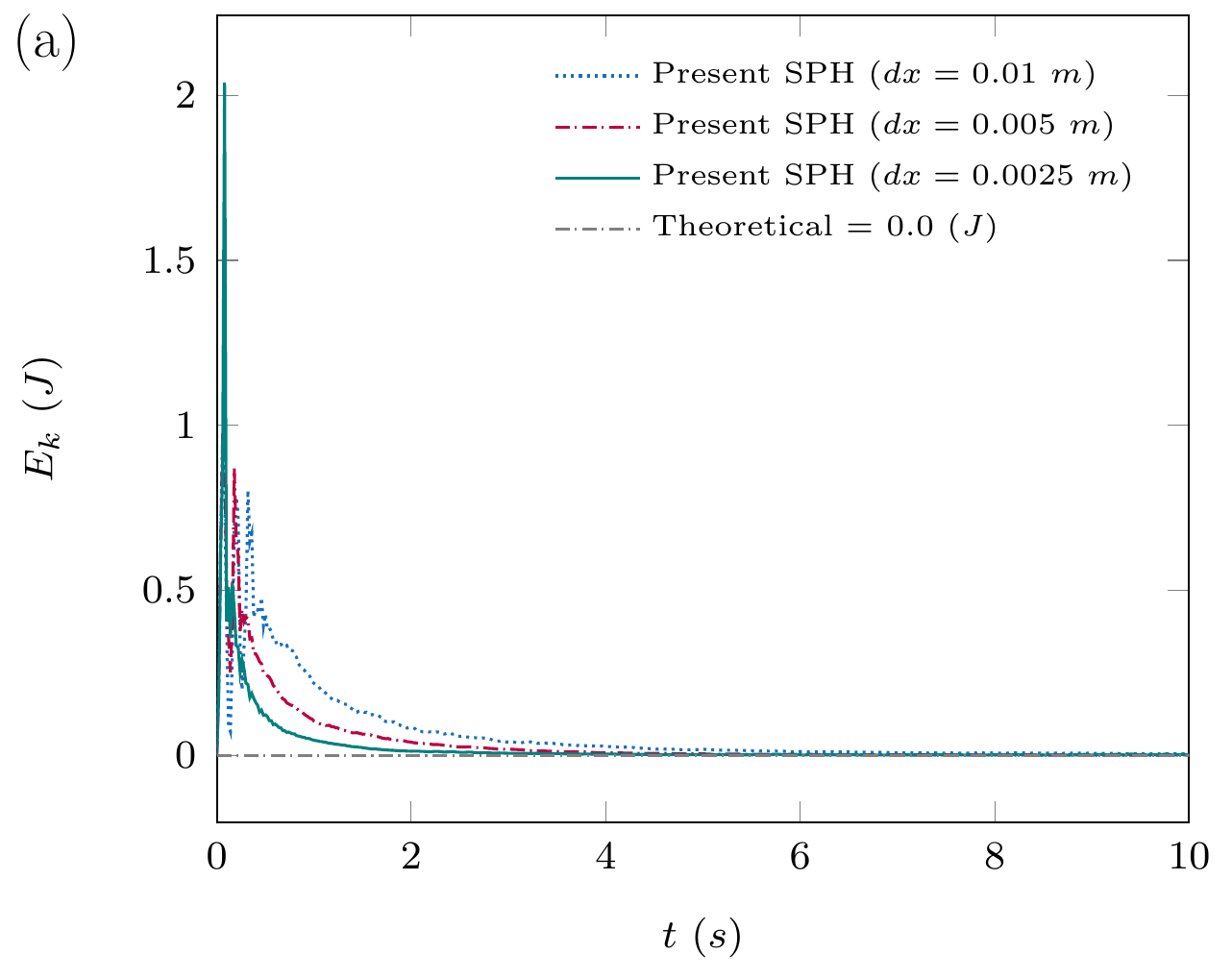}
	\hspace{0.3cm}
	\includegraphics[trim = 0mm 0mm 0mm 0mm, clip,width=0.45\linewidth]{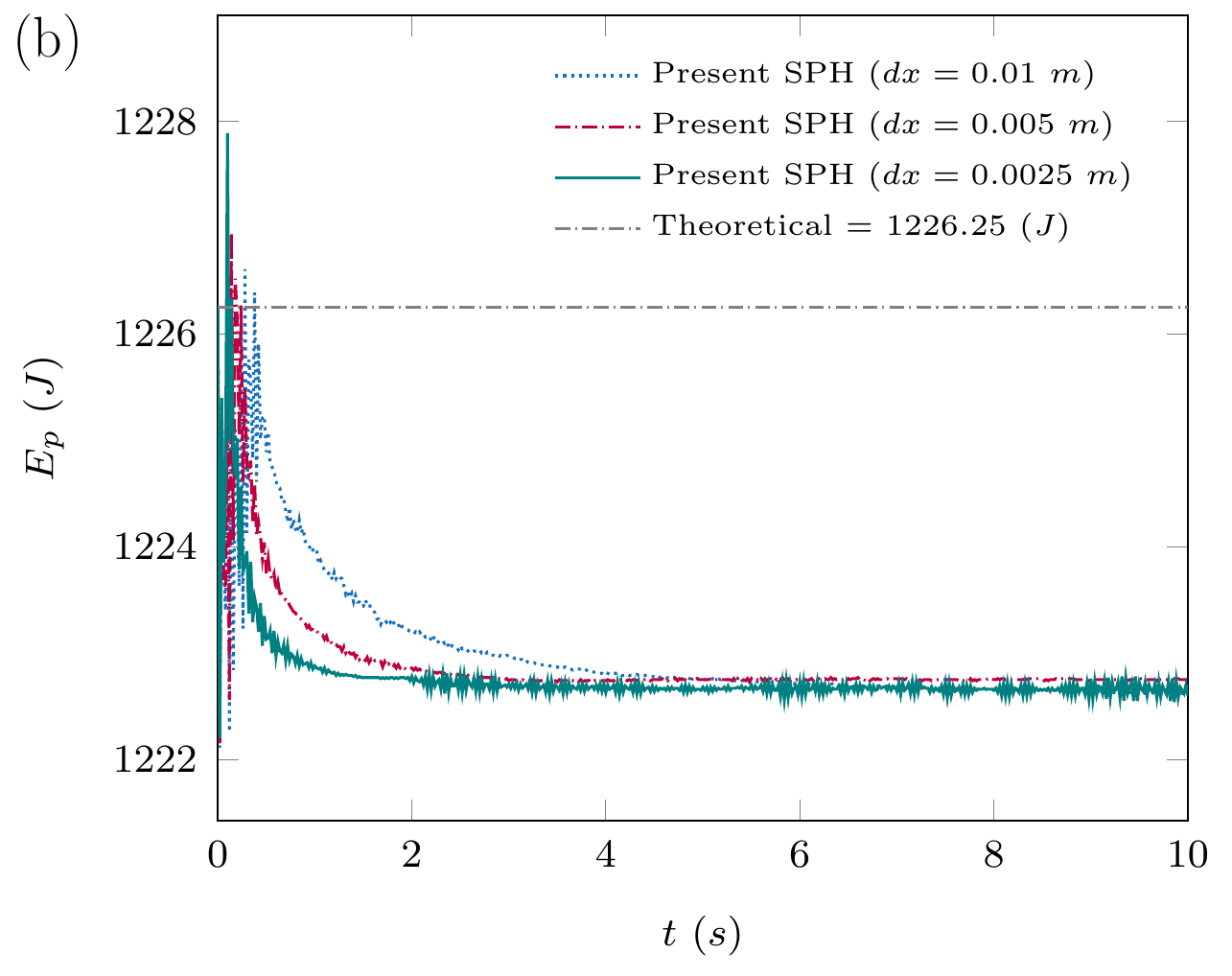}
	\caption{Two-phase hydrostatic tank: time histories of (a) global kinetic energy and (b) global potential energy of the system
		in comparison with analytical values and obtained with various particle resolutions.}
	\label{fig:staticTank_Ek_Ep}
\end{figure}

As a substantial characteristic of numerical schemes, the energy conservation properties 
of the proposed method are also investigated here. Time histories of the global kinetic and potential energies of 
the system are plotted in figure \ref{fig:staticTank_Ek_Ep}.
One can clearly observe that after the initial effects of gravity, the energy variation vanishes
and the system reaches an steady state. According to the results obtained with various particle 
resolutions, the solution evidently converges towards more accuracy with increasing resolution,
however, the increased resolution of $dx=0.0025~$m has led to considerable noises in potential energy plot,
which are due to the large number of particles contributing in the calculation of the $E_p$ values.
Considering the obtained rest situation and having the theoretical amount of energy ($T_{tot}=1226.25~$J) in mind,
it is concluded that the proposed method has dissipated only $0.26\%$ of the total energy until around $t=2~$s
and the total energy stayed conserved afterwards. 

\subsection{Two-phase sloshing tank}\label{subsec:twophase_sloshing}
The long lasting challenge of a robust wall boundary for 3-D multi-phase flows 
in the context of the SPH method has been subject of research 
since decades \citep{valizadeh2015study}. Our previously demonstrated progresses
in dealing with multi-physics highly violent flows 
\citep[e.g.][]{rezavand2020weakly,zhang2017weakly,zhang2021multi,zhang2021sphinxsys} 
are further enhanced in the present study and in this section we examine the robustness and accuracy
of the proposed method in one of the most challenging violent and complex multi-phase flows. 
If the motion frequency in a liquid sloshing tank is close to the natural 
frequency of the liquid due to the gravity wave, 
resonant condition happens, which may cause enormous impact loads on the structures.  
In this cases, where particles of a light phase are interacting with the 
wall boundaries and simultaneously with the heavy phase particles, 
some particles of the light phase penetrate into the wall and ultimately leave the 
3-D computational domain. In this section, we also demonstrate the ability of 
the proposed method to effectively prevent this phenomena and compare the
results with the method of \cite{ADAMI2012wall}.

Due to the highly non-linear phenomena in sloshing, 
similar benchmark cases have been used to validate several multi-physics SPH formulations 
\citep[see e.g.][]{khayyer2021multi,lyu2021further,zhang2020dual,gotoh2014enhancement}. 
\citet{cercos2015aquagpusph} and \citet{WINKLER2017165} have also solved this problem 
on GPUs.

\begin{figure}
	\centering
	\includegraphics[width=0.5\linewidth]{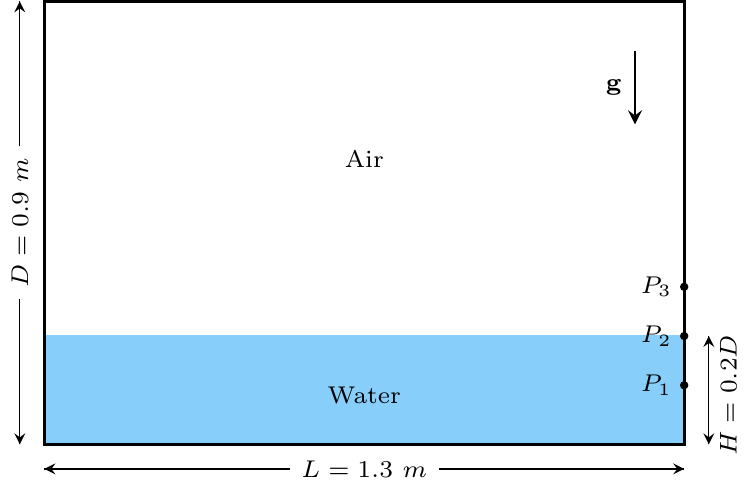}
	\caption{3-D two-phase liquid-gas sloshing tank: schematic illustration.}
	\label{fig:2phase_sloshing_config}
\end{figure}

\begin{figure}
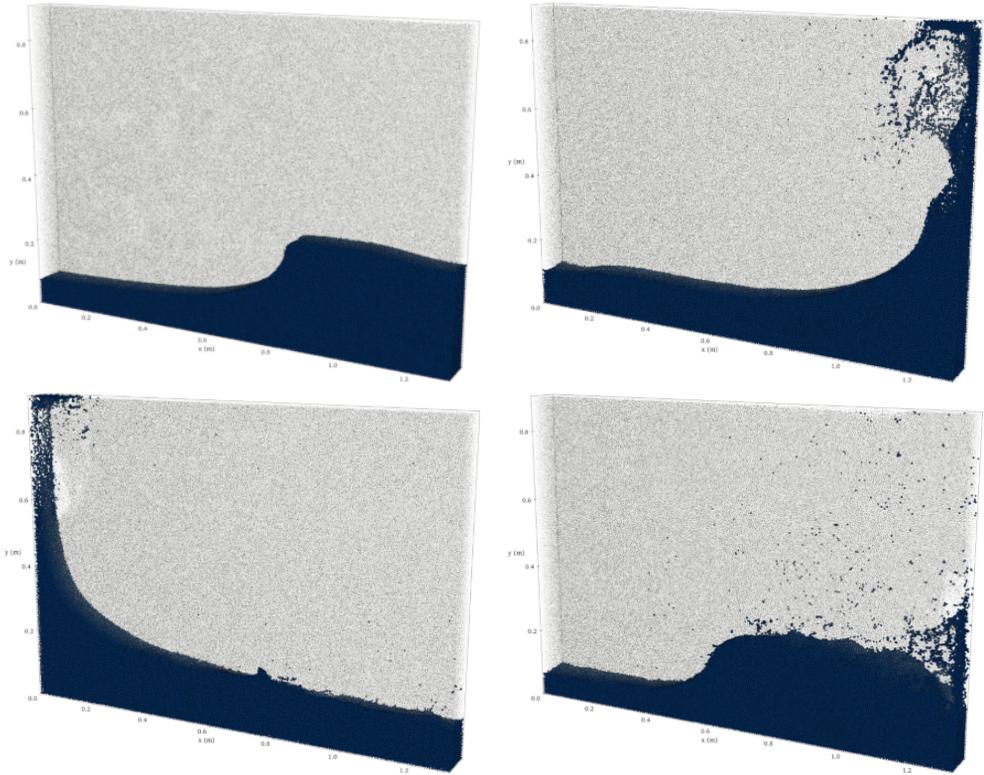

	\centering	
	\includegraphics[trim = 0mm 0mm 0mm 0mm, clip,width=0.485\textwidth]{Figs/snapshots_sloshingTank/sloshingTank_dx005_t20_HQ-eps-converted-to.png}
	\includegraphics[trim = 0mm 0mm 0mm 0mm, clip,width=0.485\textwidth]{Figs/snapshots_sloshingTank/sloshingTank_dx005_t38_HQ-eps-converted-to.png}\\
	\includegraphics[trim = 0mm 0mm 0mm 0mm, clip,width=0.485\textwidth]{Figs/snapshots_sloshingTank/sloshingTank_dx005_t47_HQ-eps-converted-to.png}
	\includegraphics[trim = 0mm 0mm 0mm 0mm, clip,width=0.485\textwidth]{Figs/snapshots_sloshingTank/sloshingTank_dx005_t625_HQ-eps-converted-to.png}
	\caption{3-D two-phase liquid-gas sloshing tank: snapshots of the flow evolution together with the details of multi-phase interface.
		The snapshots illustrate the time frames of $t=2.0,~3.8,~4.7$ and $6.25$ s and $dx = 0.005~$m.}
	\label{figs:sloshingSnapshots}
\end{figure}

In this section we consider a 3-D two-phase liquid-gas sloshing case, 
which has been experimentally studied by Rafiee et al. \citep{rafiee2011study}. 
The schematic of this problem is illustrated in figure \ref{fig:2phase_sloshing_config}. 
A tank of height $D=0.9~$m and width $L=1.3~$m 
is partially filled with water to 20\% of the height, 
i.e. $H=0.18~$m, 
while the remainder is filled by air. 
The flow is considered to be inviscid also in this case, 
where the density of water and air are set to 
$\rho_w = 1000~\mathrm{kg\cdot m^{-3}}$ and $\rho_a = 1~\mathrm{kg \cdot m^{-3}}$, respectively. 
The tank motion is defined by a sinusoidal excitation of $x = A_0sin(2.0 f_0 \pi t)$, 
where $A_0 = 0.1~$m and $f_0 = 0.496~\mathrm{s^{-1}}$ are the amplitude 
and frequency, respectively. The simulations of this section have been carried out with $dx=0.005$ m, which 
results in $1~410~752$ total particles.
Figure \ref{figs:sloshingSnapshots} shows the sloshing flow evolution within the 3-D domain
in different time frames. 
The particles are identified with their phase index and 
it can be seen that the multi-phase interface is well captured
and the dynamic nature of the flow along with the splashing phenomena are well reproduces.
Moreover, the interaction of fluid particles with the wall boundaries is efficiently 
simulated and no non-physical penetration is noted. 

\begin{figure}
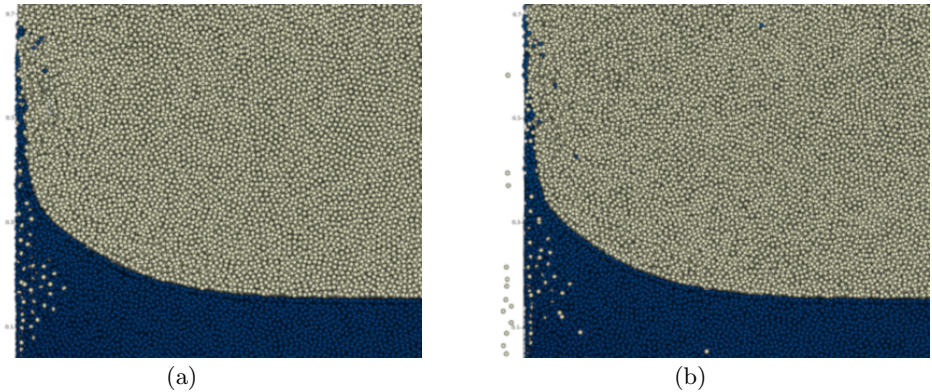

	\centering
	\subfigure []{\includegraphics[width=0.46\linewidth]{Figs/snapshots_sloshingTank/comparison_withWithoutPenalty/interface_withPenalty_t27-eps-converted-to.png}} 
	\hspace{0.3cm}
	\subfigure []{\includegraphics[width=0.46\linewidth]{Figs/snapshots_sloshingTank/comparison_withWithoutPenalty/interface_withoutPenalty_t27-eps-converted-to.png}}
	\caption{3-D two-phase liquid-gas sloshing tank: zoom-in views of the flow simulations (a) with and (b) without
		the $\mathbf{F}_p$ penalty force term being applied between particles of the light phase and the wall boundary 
		particles.}
	\label{figs:sloshingPenetraionZoomInSnapshots}
\end{figure}

The efficiency of the proposed method in preventing the particle penetration
phenomena mentioned in previous sections is detailed in figure \ref{figs:sloshingPenetraionZoomInSnapshots}.
It can be clearly observed that without applying the $\mathbf{F}_p$
penalty force term between particles of the light phase and the wall boundary 
particles, several particles of the light phase have penetrated into the 
wall boundaries at a triple point where all three phases, namely water, air and solid walls, are present.
This behaviour continues to the next time steps and the particles ultimately leave the 
computational domain, which violates the conservation of mass principle and 
causes numerical instability. The proposed method effectively prevents this non-physical 
behaviour and leads to accurate and stable solutions.

\begin{figure}
	\centering
	\subfigure []{\includegraphics[width=0.46\linewidth]{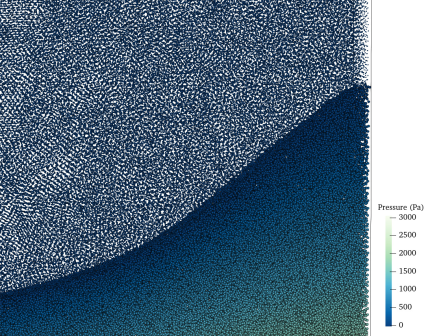}} 
	\hspace{0.3cm}
	\subfigure []{\includegraphics[width=0.46\linewidth]{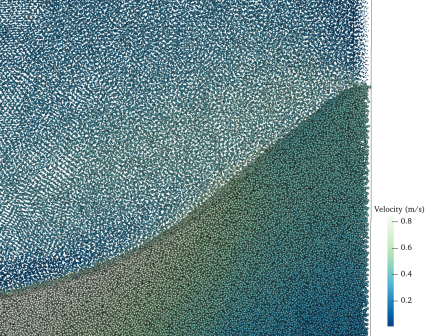}}
	\caption{3-D two-phase liquid-gas sloshing tank: zoom-in views of the (a) pressure field and (b) velocity
		field at a triple point where liquid, gas and solid wall particles interact at $t=1.75$ s
		where the first strong impact occurs.}
	\label{figs:sloshingZoomInSnapshots}
\end{figure}

In order to further analyse the 
accuracy and performance of the proposed wall boundary condition in some 
cases where triple points occur, the interaction of 
fluid particles with wall boundaries at a triple point
is shown in figure \ref{figs:sloshingZoomInSnapshots}.
It can be evidently seen that the interactions of both liquid and gas particles with 
the wall particles are very well reproduced, 
the interfaces between various phases
are sharply maintained and quite smooth pressure and velocity fields are achieved.
We can see that with the herein proposed wall boundary treatment an accurate and efficient
interaction between fluid and wall particles is realised and non-physical penetration is prevented. 
Note that as the particles
of the wall boundaries are constantly fixed, for easier interpretations of the figures we have not depicted them here.

\begin{figure}
	\centering	
	\includegraphics[trim = 0mm 0mm 0mm 0mm, clip,width=0.485\textwidth]{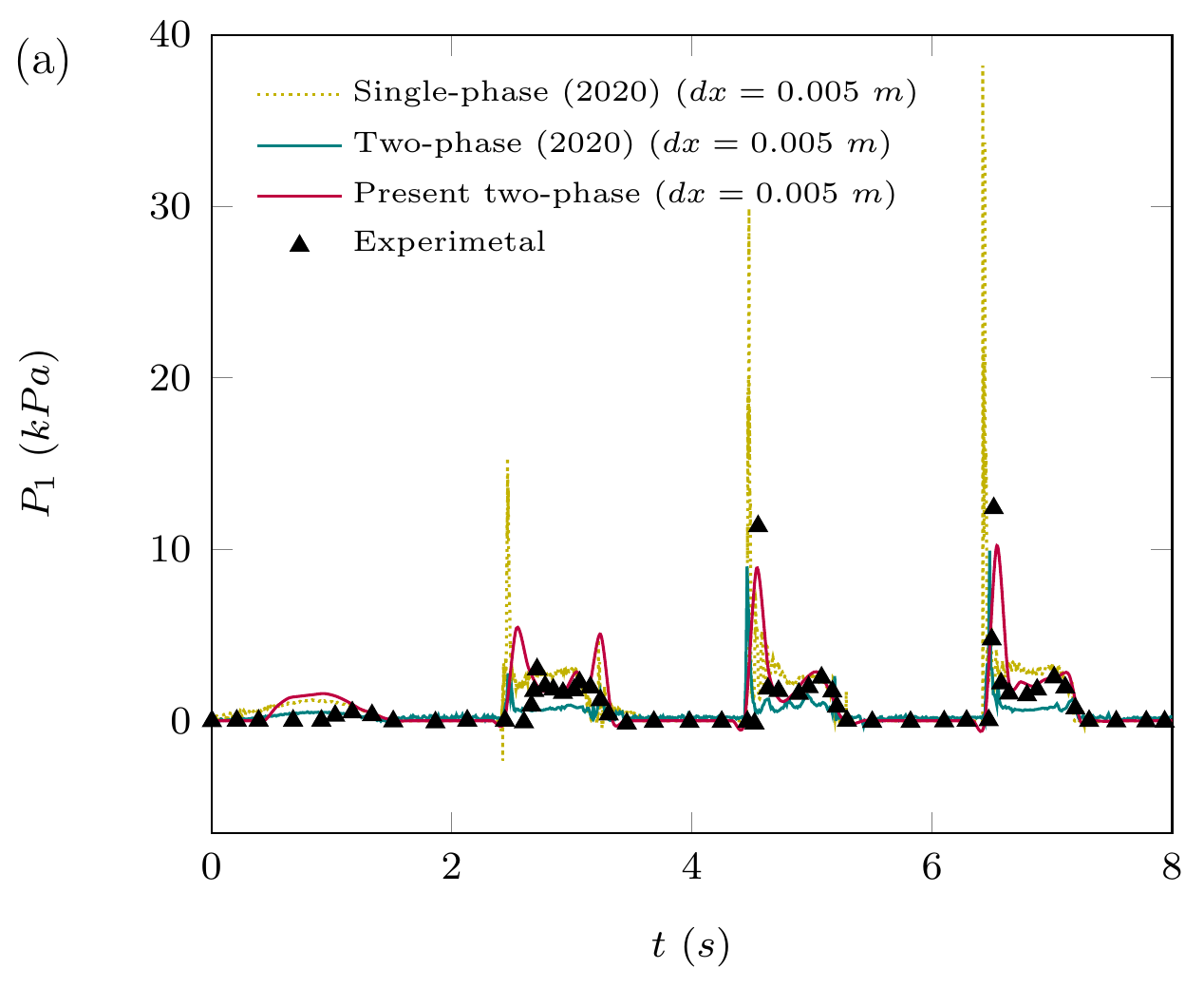}
	\includegraphics[trim = 0mm 0mm 0mm 0mm, clip,width=0.485\textwidth]{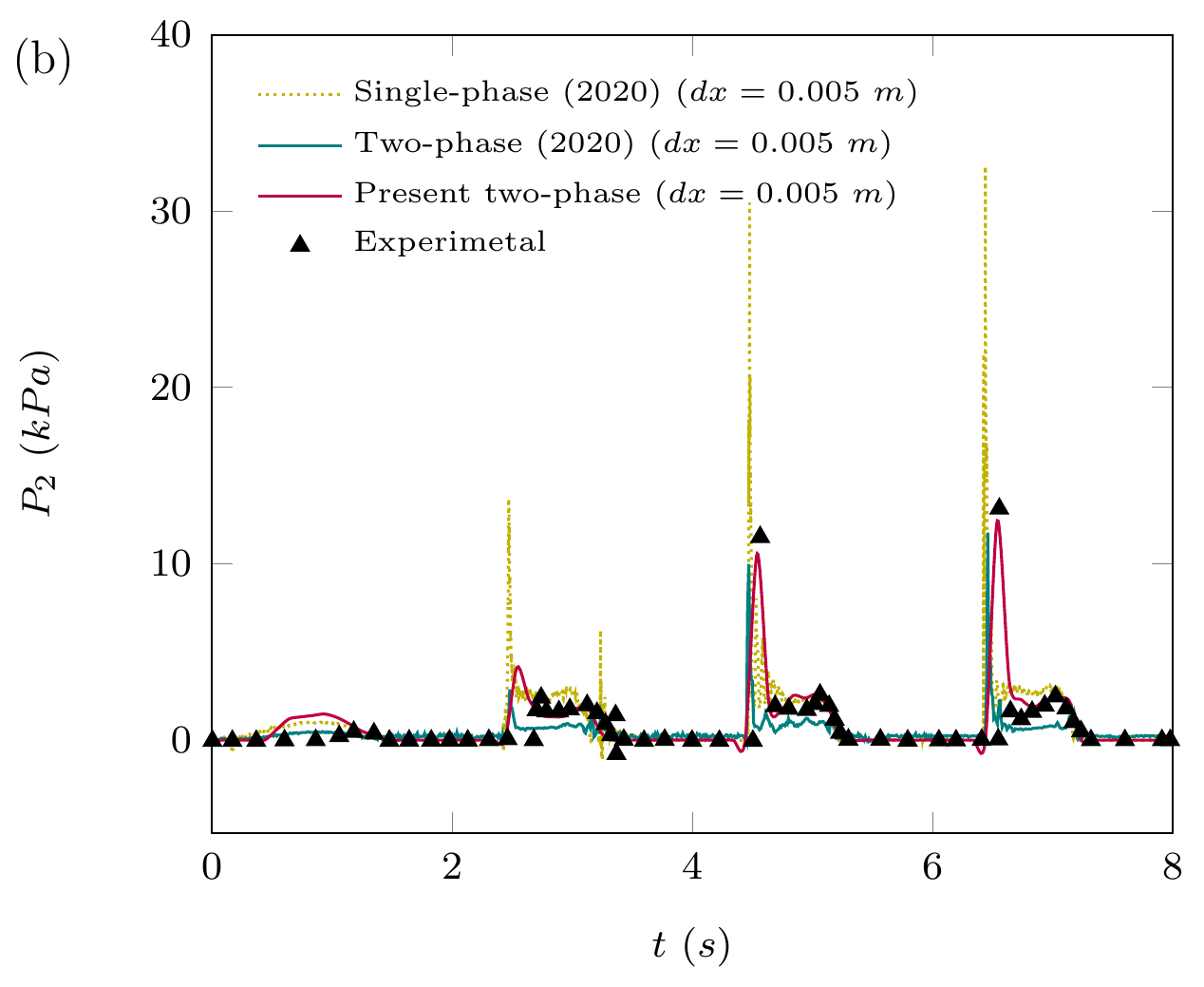}\\
	\includegraphics[trim = 0mm 0mm 0mm 0mm, clip,width=0.485\textwidth]{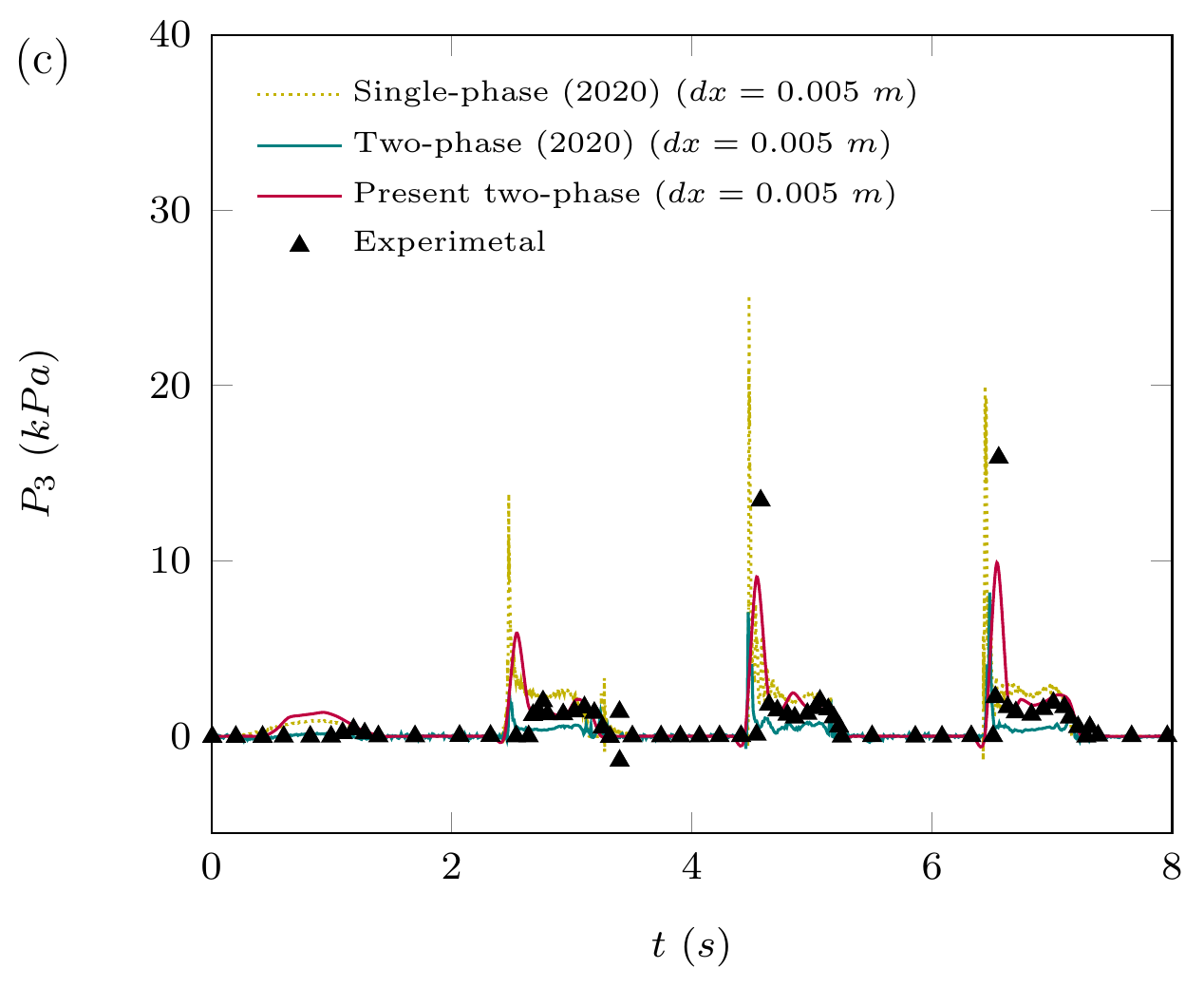}
	\caption{3-D two-phase liquid-gas sloshing tank: time histories of the impact pressure at 
		(a) $P_1$, (b) $P_2$ and (c) $P_3$ sensors
		in comparison with the experimental results of \citet{rafiee2011study}, previous two-phase SPH simulations of 
		\citet{rezavand2020weakly} and single-phase SPH simulations of \citet{zhang2020dual}}
	\label{figs:sloshing_impactPressureSignals}
\end{figure}

The obtained results for the 3-D two-phase liquid-gas sloshing tank by the present method are
also quantitatively validated against experimental results of \citet{rafiee2011study} 
in figure \ref{figs:sloshing_impactPressureSignals}. The pressure signals at
three sensors shown in figure \ref{fig:2phase_sloshing_config} are calculated 
and plotted here. A good agreement is evidently achieved between the experiments and 
the results obtained by the presented methodology.
More importantly, the peak impact pressures predicted by the present method are 
found to be in a clearly better agreement with the experiments in comparison to the previous two-phase
and single-phase simulations of \citet{rezavand2020weakly} and \citet{zhang2020dual}, respectively.
These analysis demonstrate that the newly proposed wall boundary condition can 
outperform the previous methodologies in the simulation of highly violent impacts and
prediction of the impact pressure, while achieves also more accurate simulations.

\subsection{Dam break with an obstacle}\label{subsec:dambreak_withObstacle}
In this section we consider a 3-D dam break case with an obstacle placed on the downstream horizontal bed, 
which is known as the SPHERIC benchmark \#2 and its detailed description can be found
in \url{https://spheric-sph.org/tests/test-2}.
This test case has been studied by \citet{kleefsman2005volume} first experimentally and then by 
an Eulerian volume of fluid (VOF) method. Later, this
test case was simulated with the standard WCSPH and 
truly incompressible SPH (ISPH) methods by \citet{lee2010application} and was recently also studied using
a weakly compressible moving particle semi-implicit (WCMPS) by \citet{jandaghian2021enhanced}. 
In our simulations, particles are initially
placed regularly on a Cartesian grid and in order to further analyse the 
convergence properties of the proposed method,
we carry out the simulations with three different particle resolutions, viz. 
$dx=0.032, ~0.016$ and $0.008~$m, which correspond to $69~773$, $349~340$ and $2~061~483$ number of particles, respectively.  

\begin{figure}
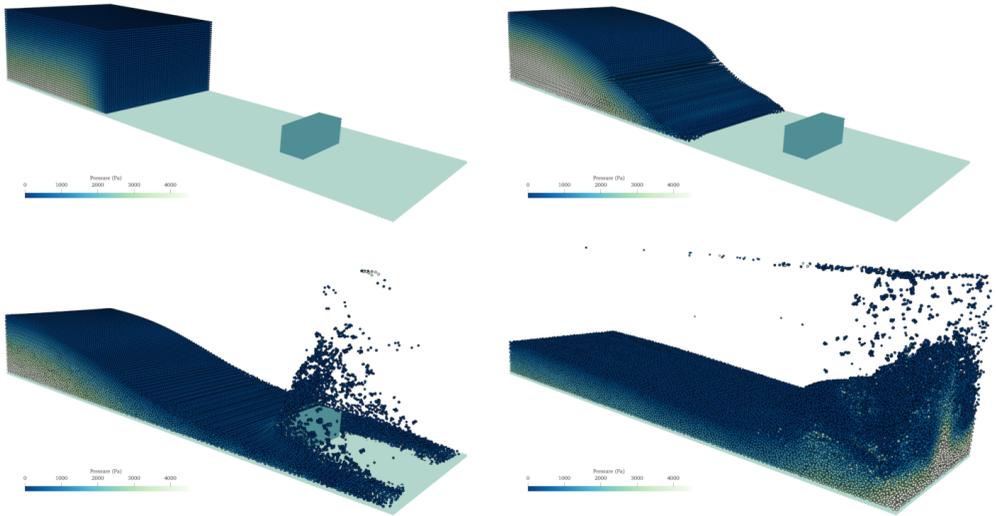

	\centering	
	\includegraphics[trim = 0mm 0mm 0mm 0mm, clip,width=0.485\textwidth]{Figs/snapshots_damObstacle/pressure_snapshot_t0_HQ-eps-converted-to.png}
	\includegraphics[trim = 0mm 0mm 0mm 0mm, clip,width=0.485\textwidth]{Figs/snapshots_damObstacle/pressure_snapshot_t30_HQ-eps-converted-to.png}\\
	\includegraphics[trim = 0mm 0mm 0mm 0mm, clip,width=0.485\textwidth]{Figs/snapshots_damObstacle/pressure_snapshot_t60_HQ-eps-converted-to.png}
	\includegraphics[trim = 0mm 0mm 0mm 0mm, clip,width=0.485\textwidth]{Figs/snapshots_damObstacle/pressure_snapshot_t120_HQ-eps-converted-to.png}
	\caption{3-D dam break with obstacle: 
		snapshots of the flow evolution together with pressure distribution within the domain.
		The snapshots illustrate the time frames of $t=0,~0.3,~0.6$ and $1.2$ s and $dx = 0.016~$m.}
	\label{figs:dambreakSnapshots_pressure}
\end{figure}

\begin{figure}
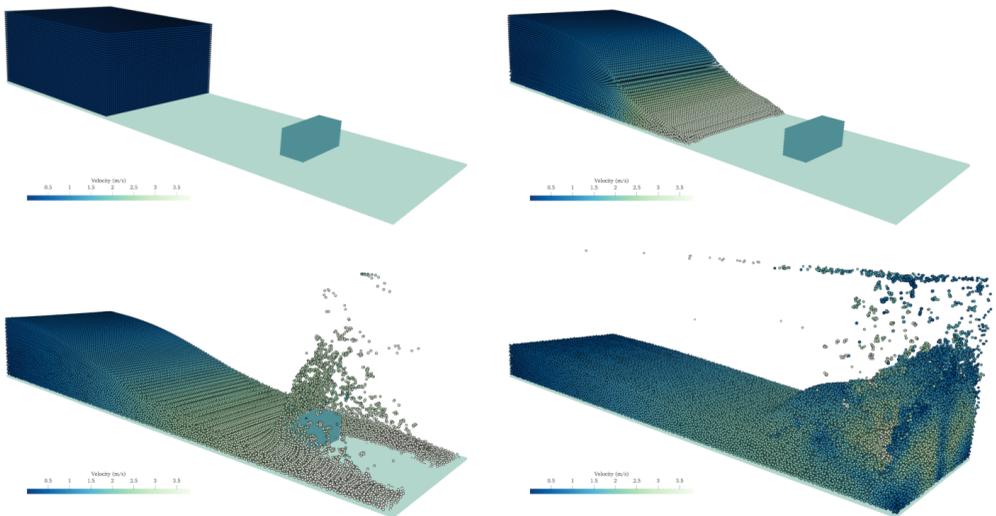

	\centering	
	\includegraphics[trim = 0mm 0mm 0mm 0mm, clip,width=0.485\textwidth]{Figs/snapshots_damObstacle/velocity_snapshot_t0_HQ-eps-converted-to.png}
	\includegraphics[trim = 0mm 0mm 0mm 0mm, clip,width=0.485\textwidth]{Figs/snapshots_damObstacle/velocity_snapshot_t30_HQ-eps-converted-to.png}\\
	\includegraphics[trim = 0mm 0mm 0mm 0mm, clip,width=0.485\textwidth]{Figs/snapshots_damObstacle/velocity_snapshot_t60_HQ-eps-converted-to.png}
	\includegraphics[trim = 0mm 0mm 0mm 0mm, clip,width=0.485\textwidth]{Figs/snapshots_damObstacle/velocity_snapshot_t120_HQ-eps-converted-to.png}
	\caption{3-D dam break with obstacle: 
		snapshots of the flow evolution together with velocity distribution within the domain.
		The snapshots illustrate the time frames of $t=0,~0.3,~0.6$ and $1.2$ s and $dx = 0.016~$m.}
	\label{figs:dambreakSnapshots_velocity}
\end{figure}

The evolution of the breaking dam flow is illustrated in several snapshots in figures \ref{figs:dambreakSnapshots_pressure} 
and \ref{figs:dambreakSnapshots_velocity}, where the contours of pressure and velocity fields are shown, respectively.
All physical features of the flow are very well captured and the distribution of both pressure and velocity fields are
quite smooth at the vicinity of the solid walls, as well as the free-surface. The interaction of 
the water flow with the obstacle at $t=0.6~$s is also very well reproduced and the splash-up 
after the surge wave impacts the obstacle is well exhibited. The above observations demonstrate
the effectiveness and accuracy of the proposed wall boundary treatment.

\begin{figure}
	\centering	
	\includegraphics[trim = 0mm 0mm 0mm 0mm, clip,width=0.485\textwidth]{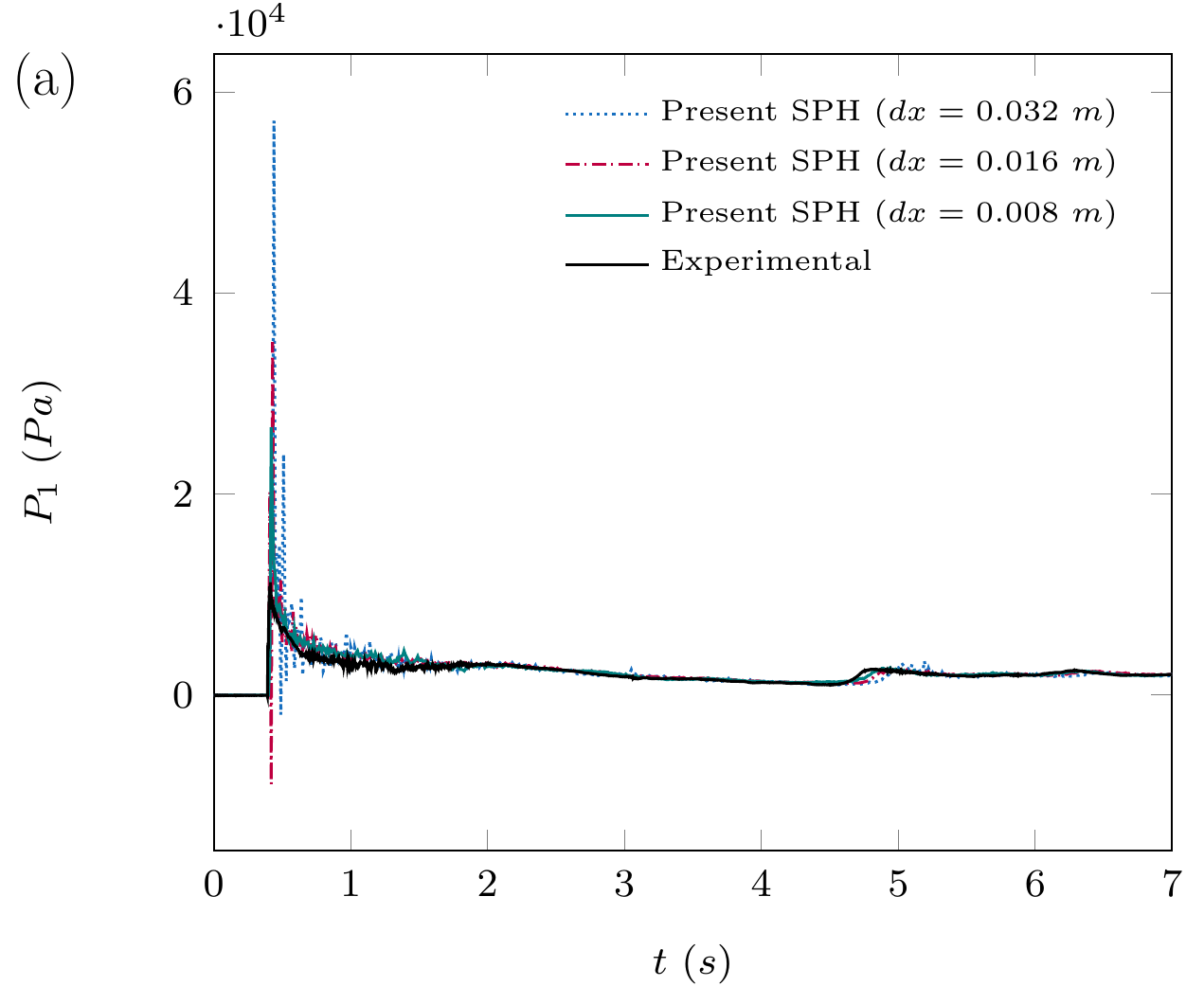}
	\includegraphics[trim = 0mm 0mm 0mm 0mm, clip,width=0.485\textwidth]{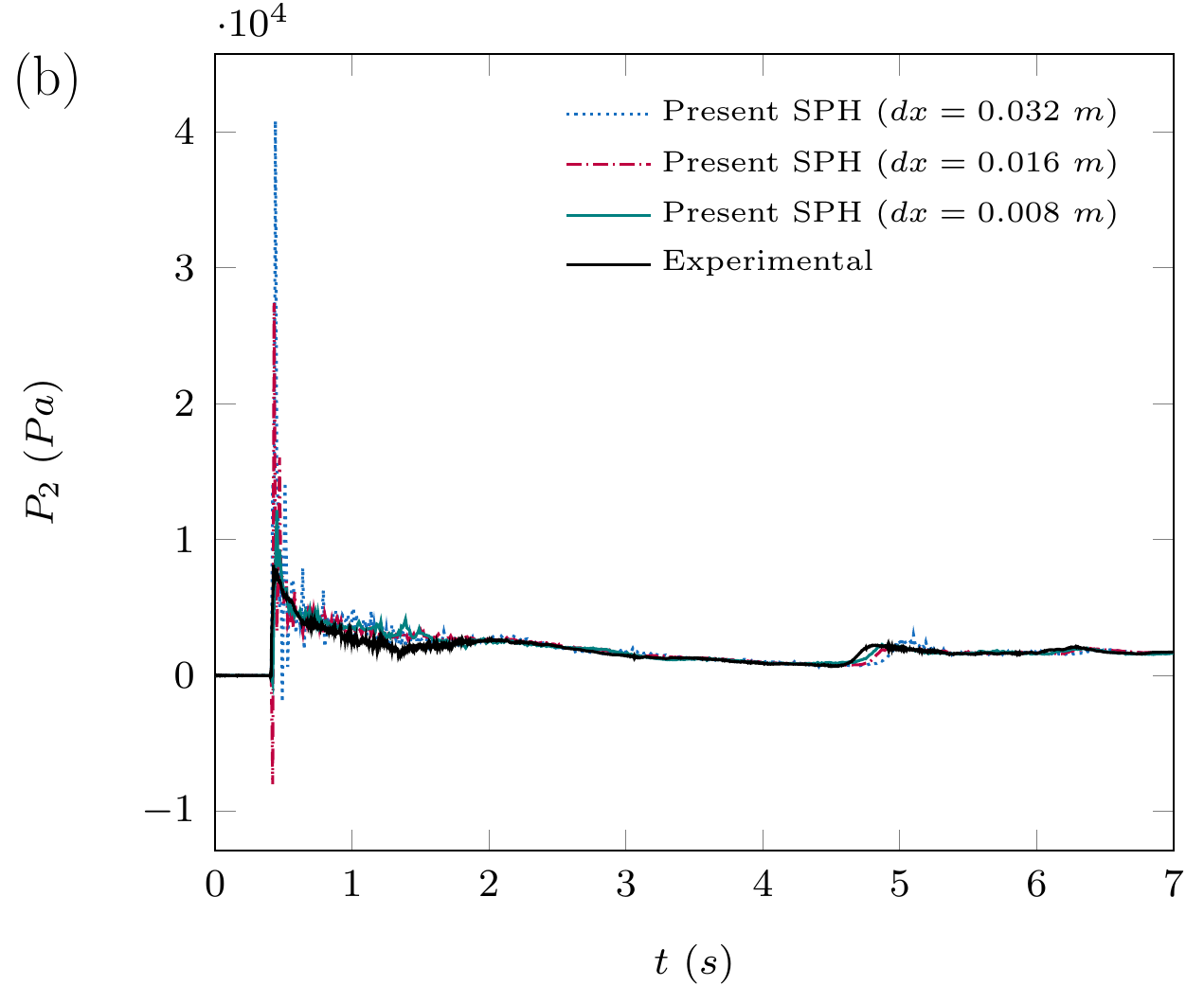}\\
	\includegraphics[trim = 0mm 0mm 0mm 0mm, clip,width=0.485\textwidth]{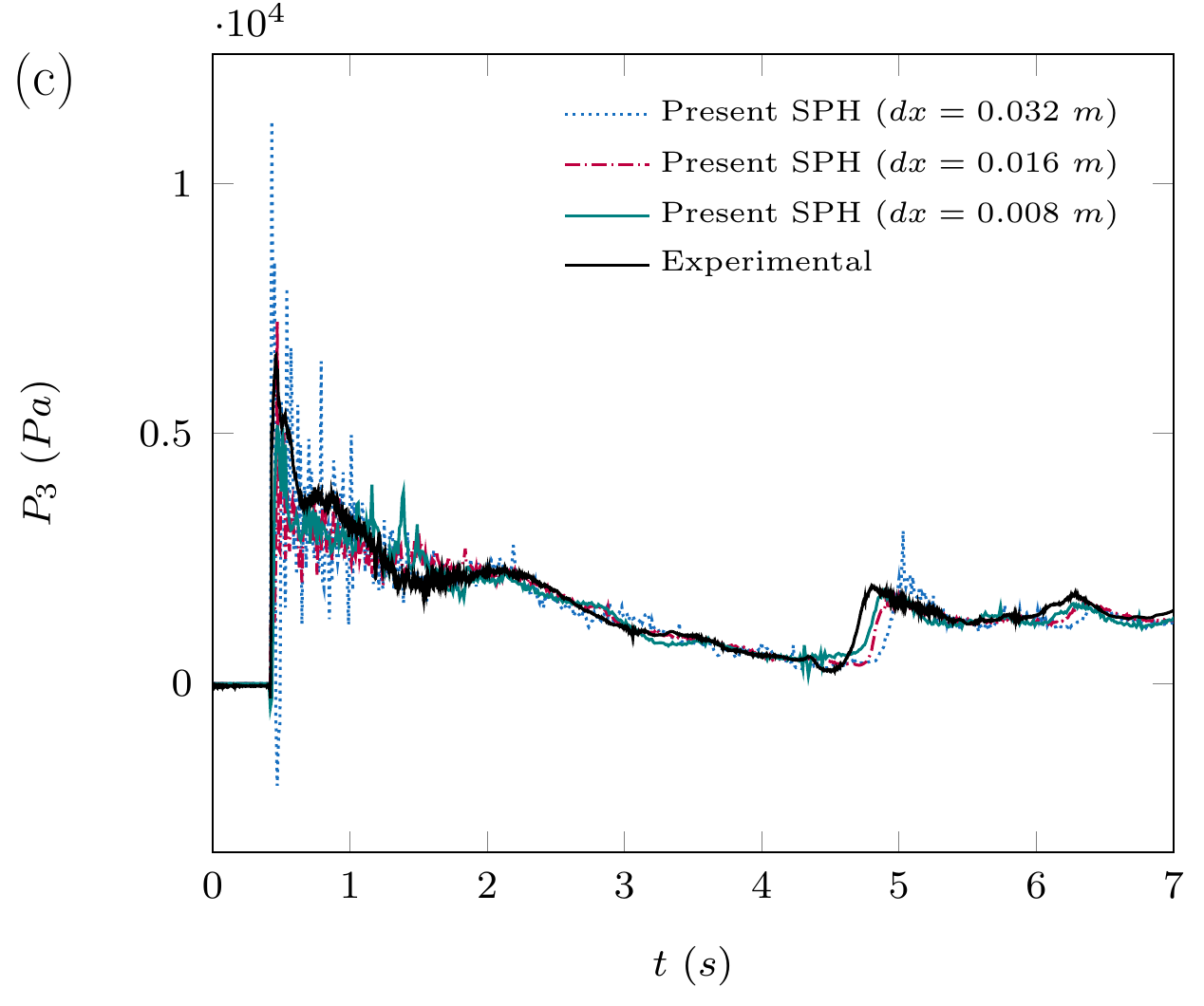}
	\includegraphics[trim = 0mm 0mm 0mm 0mm, clip,width=0.485\textwidth]{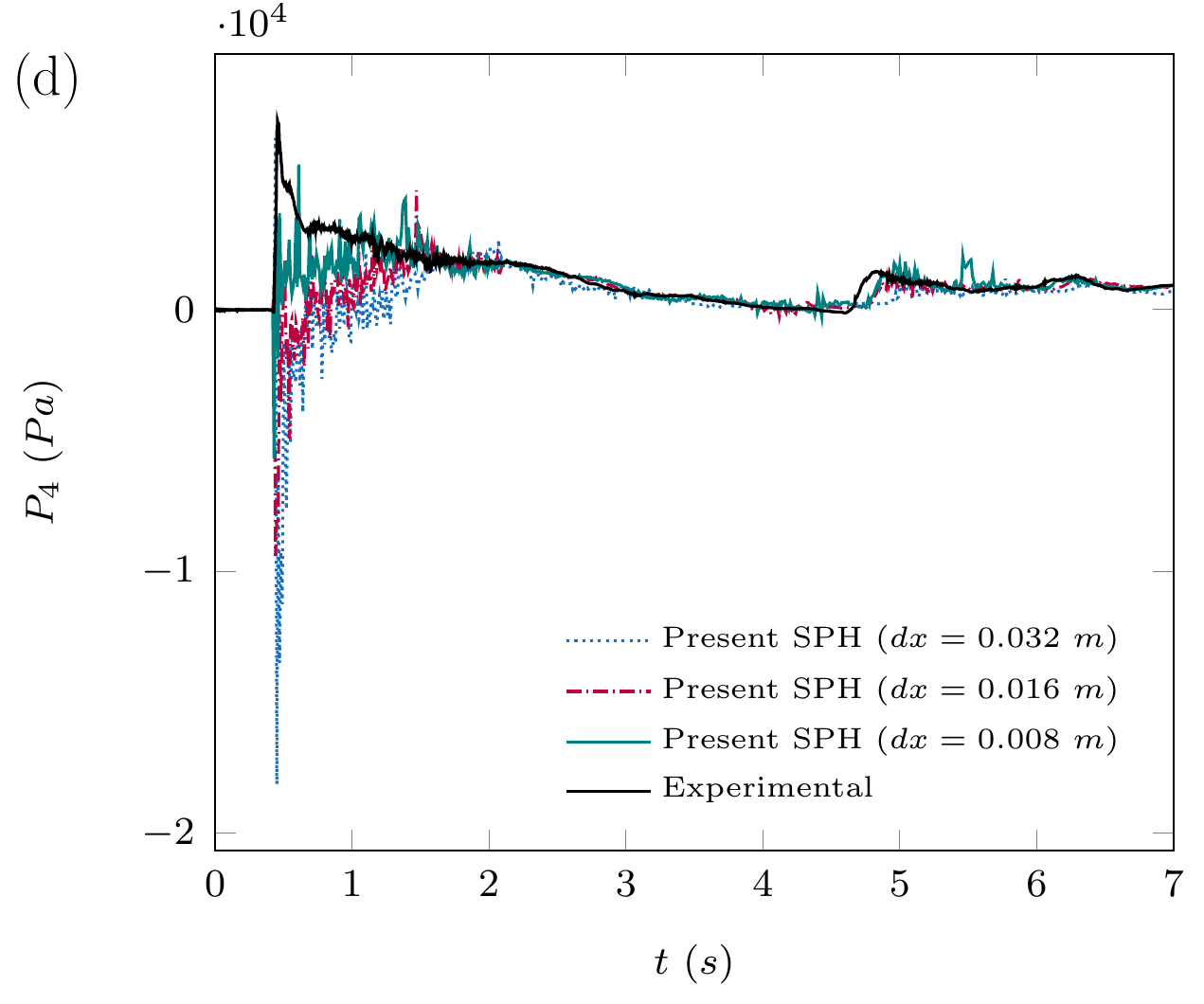}
	\caption{3-D dam break with obstacle: 
		time histories of the impact pressure at (a) $P_1$, (b) $P_2$, (c) $P_3$ and (d) $P_4$ sensors
		with different particle resolutions and	in comparison with the experimental results of \citet{kleefsman2005volume}.}
	\label{figs:dambreak_impactPressureSignals}
\end{figure}

In order to further validate the accuracy of the proposed method,
we quantitatively compare the values of impact pressure ($P_i,~i=1,\,\ldots,\,4$) and wave height 
($H_i,~i\in\{2,3,4\}$) obtained from seven sensors as describes in the test case documentation 
with the same names. The measured pressure signals are obtained by the method explained by
\citet{zhang2017weakly}. The pressure signals cause by the violent impact of 
the generated water wave are compared against experiments of \citet{kleefsman2005volume} 
in figure \ref{figs:dambreak_impactPressureSignals}.
A good agreement is noted between the present SPH simulations and the experimental results.
The peak impact pressure signals are well captured and also the long-term behaviours of the residual pressures are
well reproduced. The pressure peak at sensor $P_4$ is slightly underestimated by the proposed method, however, the 
calculated pressure signals converge to the experimental results as the 
particle resolution increases. The sensor $P_4$ is located at the top of the obstacle with the 
presence of free-surface and effects of air, which might have caused the underestimation.
A similar behaviour is however reported also by \citet{mokos2017shifting} for both higher sensors ($P_3$ and $P_4$)
although they have carried out multi-phase simulation with the consideration of the air phase.
It can be observed that our proposed wall treatment has evidently improved the pressure predictions
in comparison to their multi-phase simulation for both challenging sensors, $P_3$ and $P_4$.

\begin{figure}
	\centering	
	\includegraphics[trim = 0mm 0mm 0mm 0mm, clip,width=0.485\textwidth]{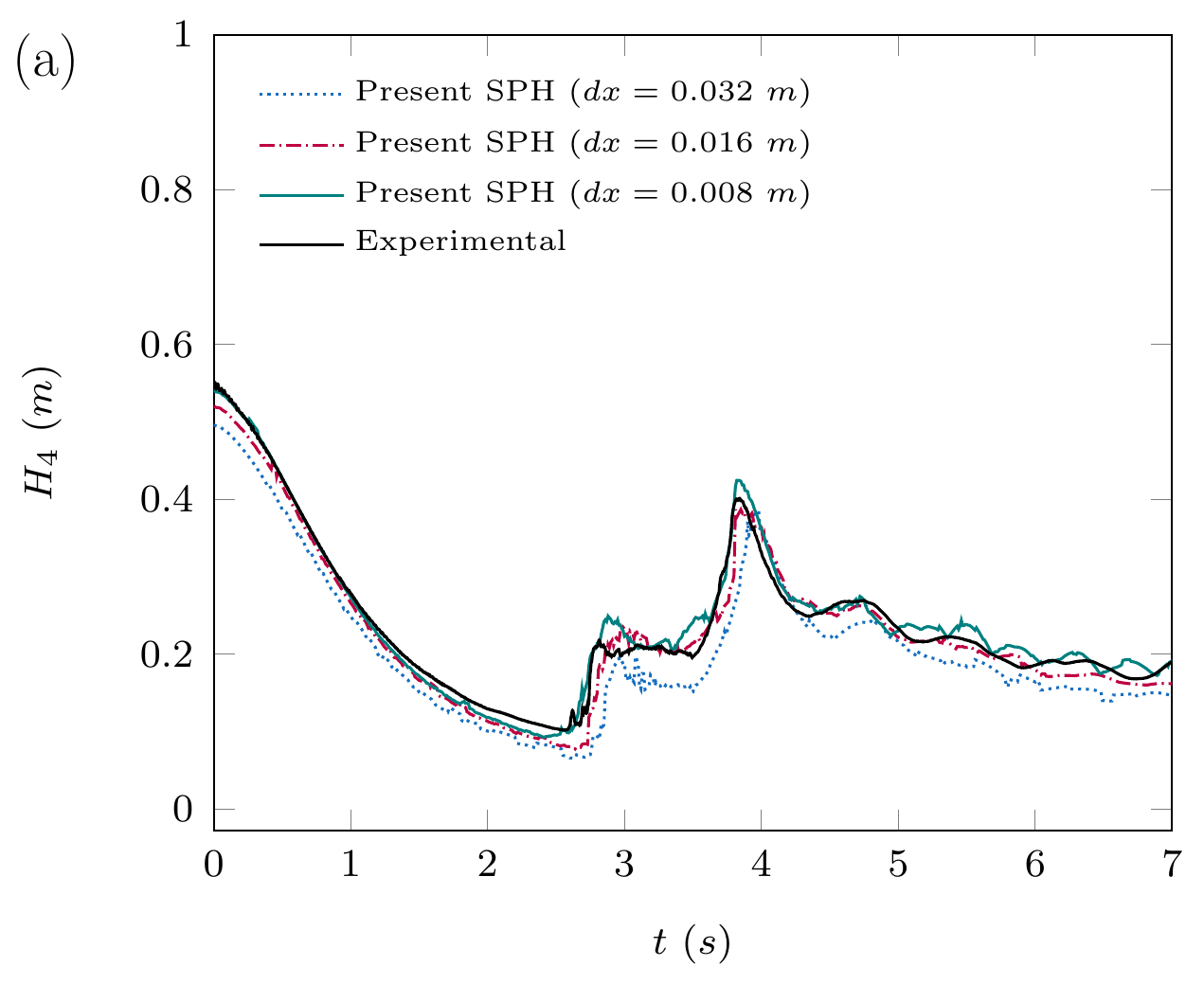}
	\includegraphics[trim = 0mm 0mm 0mm 0mm, clip,width=0.485\textwidth]{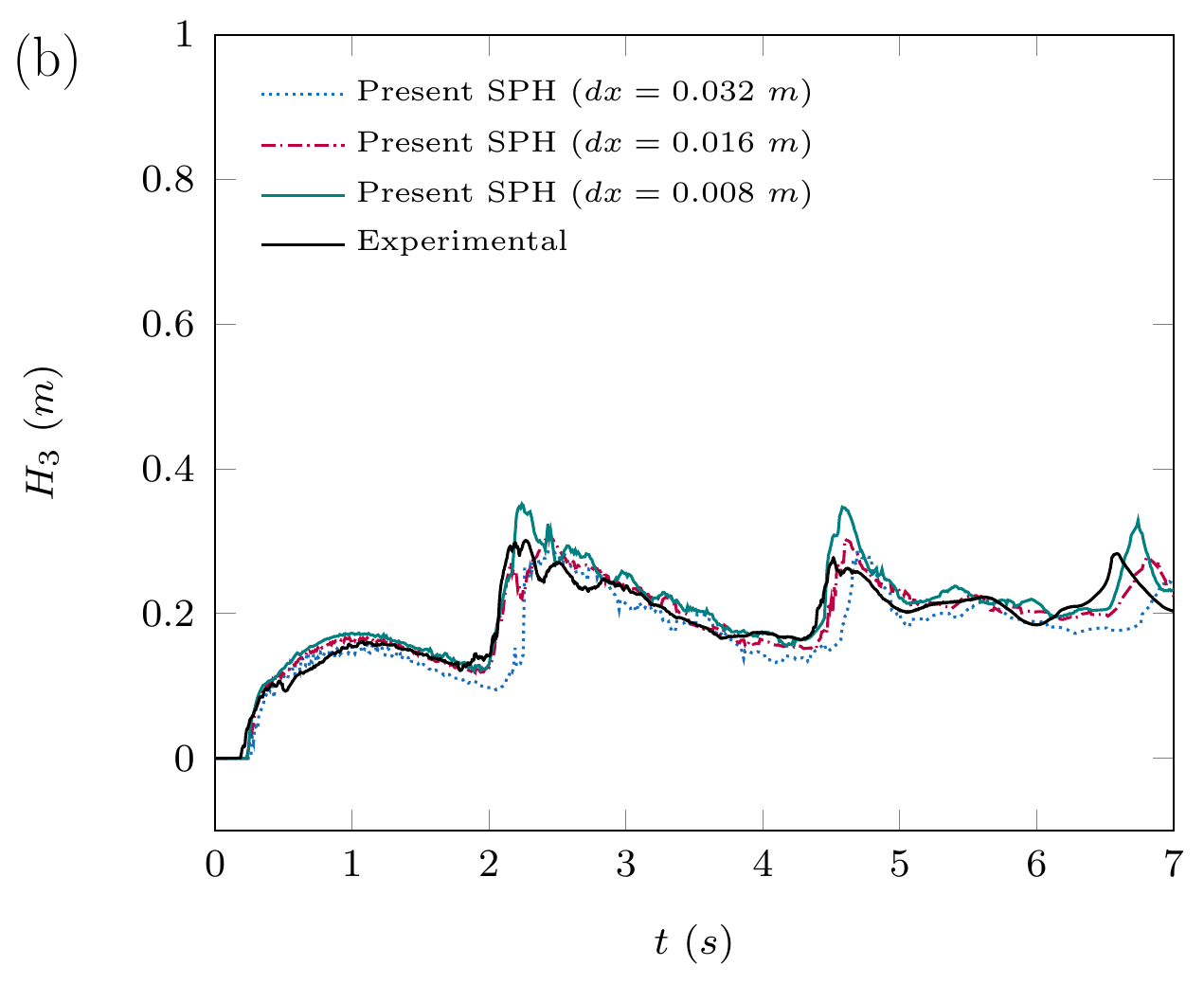}\\
	\includegraphics[trim = 0mm 0mm 0mm 0mm, clip,width=0.485\textwidth]{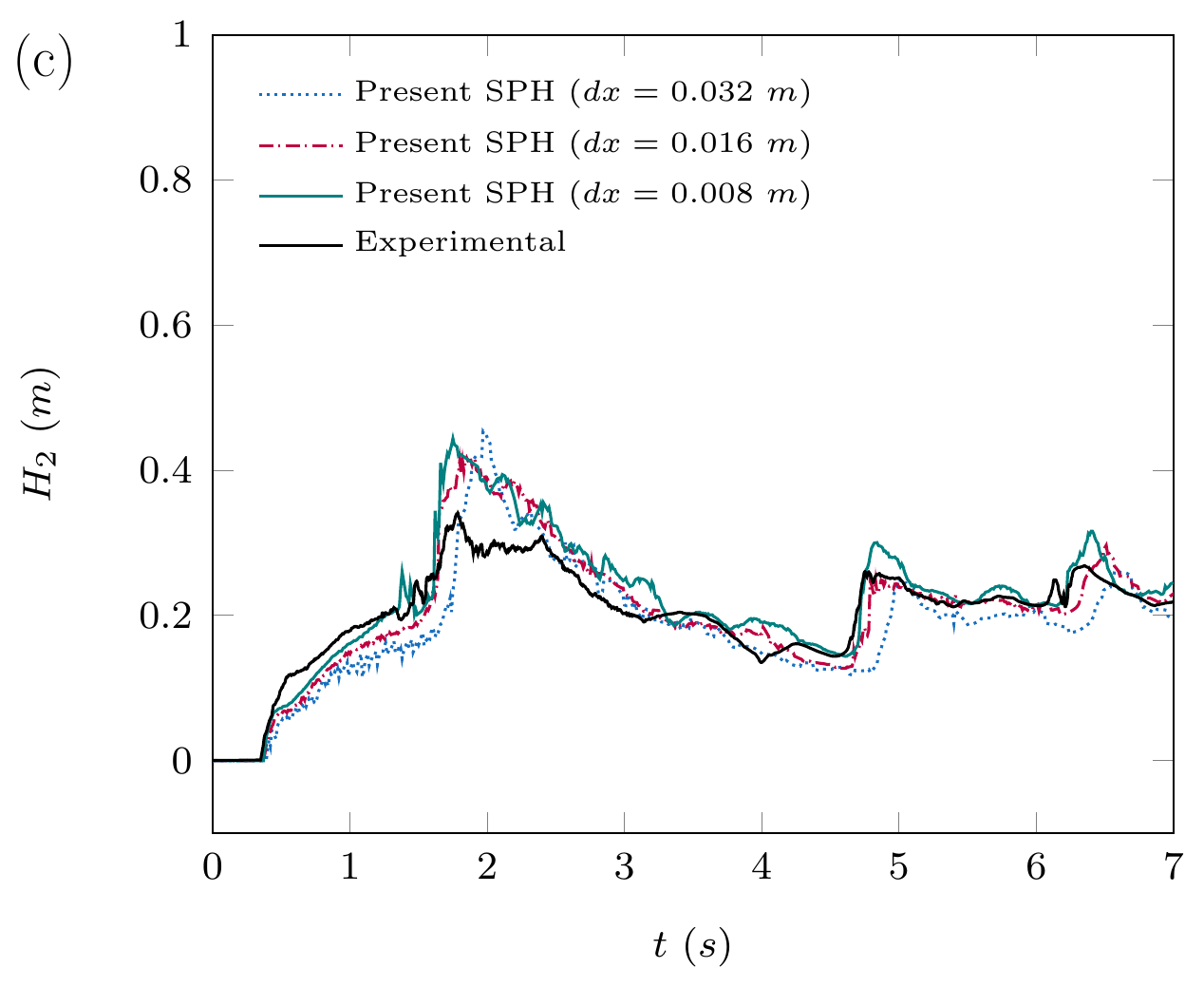}
	\caption{3-D dam break with obstacle: 
		time histories of the wave height at (a) $H_4$, (b) $H_3$ and (c) $H_2$
		with different particle resolutions and	in comparison with the experimental results of \citet{kleefsman2005volume}.}
	\label{figs:dambreak_waveHeight}
\end{figure}

Figure \ref{figs:dambreak_waveHeight} portrays the time history of the water levels recorded 
at three different probes as identified by the experiment documentation. 
A good agreement with the experiments is noted also here and the long-term behaviour of wave height is well reproduced. 
The run-up and the water level peaks are slightly overestimated, which are attributed to 
the inviscid nature of the method that, as discussed in \ref{subsec:twophase_staticTank}, introduces a marginal energy dissipation. 
It is also worth noting that in these plots the proposed method demonstrates a reasonable rate of convergence, as seen
also in previous test cases.

\subsection{Dam break with a wedge}\label{subsec:dambreak_withWedge}
\begin{figure}
	\centering
	\includegraphics[width=0.5\linewidth]{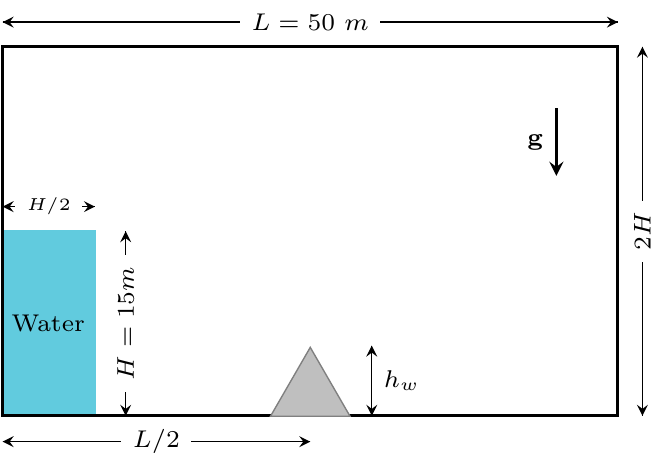}
	\caption{3-D dam break with a wedge: schematic illustration.}
	\label{fig:damBreak_wedge_config}
\end{figure}

Sharp boundaries and accurate reproduction of the resulting fluid-wall interactions
have also always been a challenge for the SPH method. 
In order to further asses the proposed method in 3-D arbitrary geometries with sharp boundaries,
here we consider a dam break flow with a wedge located at the 
middle of the downstream bed. The configuration of the problem is described in figure \ref{fig:damBreak_wedge_config}.
A similar 2-D test case has already been used by \citet{valizadeh2015study}, however, 
we consider a wedge of $\pi/3$ rad angle, which results in sharper geometries and more abrupt 
changes in the flow pattern. The height of the wedge is $h_w=5.5~$m and here we set an initial
inter-particle distance of $dx=h_w /40$ corresponding to $5~734~250$ number of particles.

\begin{figure}
	\centering	
	\includegraphics[trim = 0mm 0mm 0mm 0mm, clip,width=0.485\textwidth]{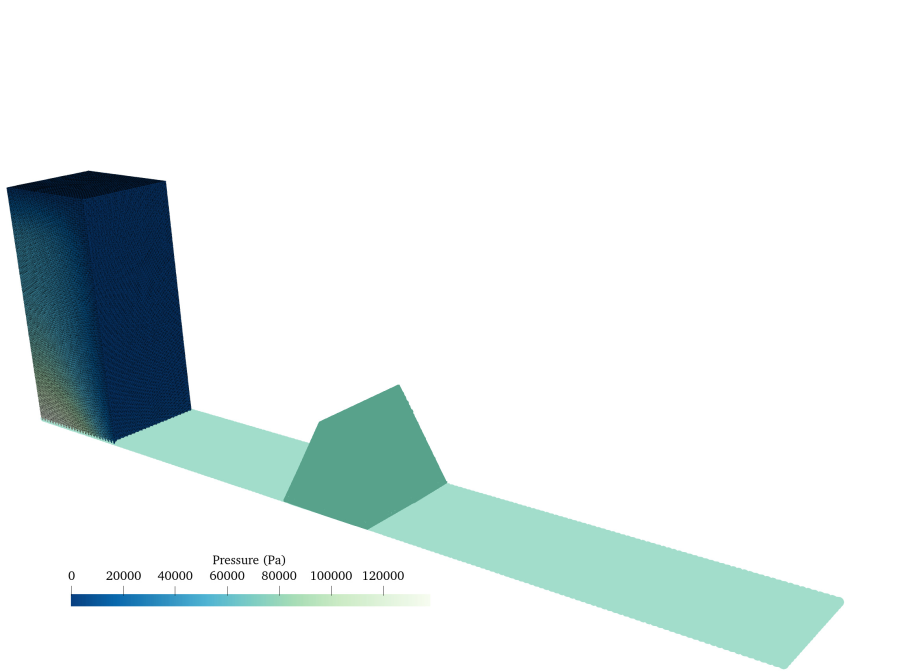}
	\includegraphics[trim = 0mm 0mm 0mm 0mm, clip,width=0.485\textwidth]{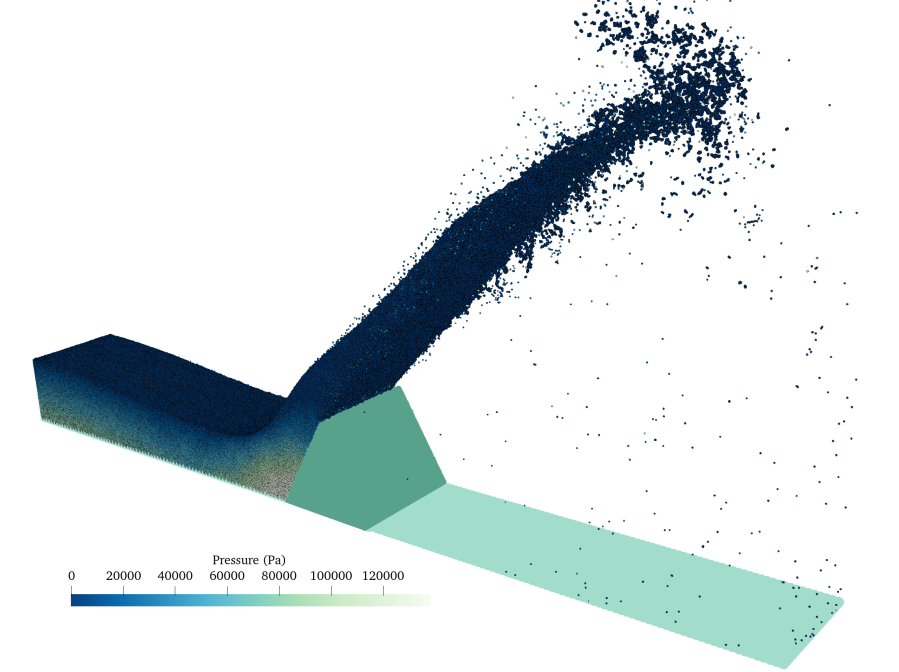}\\
	\includegraphics[trim = 0mm 0mm 0mm 0mm, clip,width=0.485\textwidth]{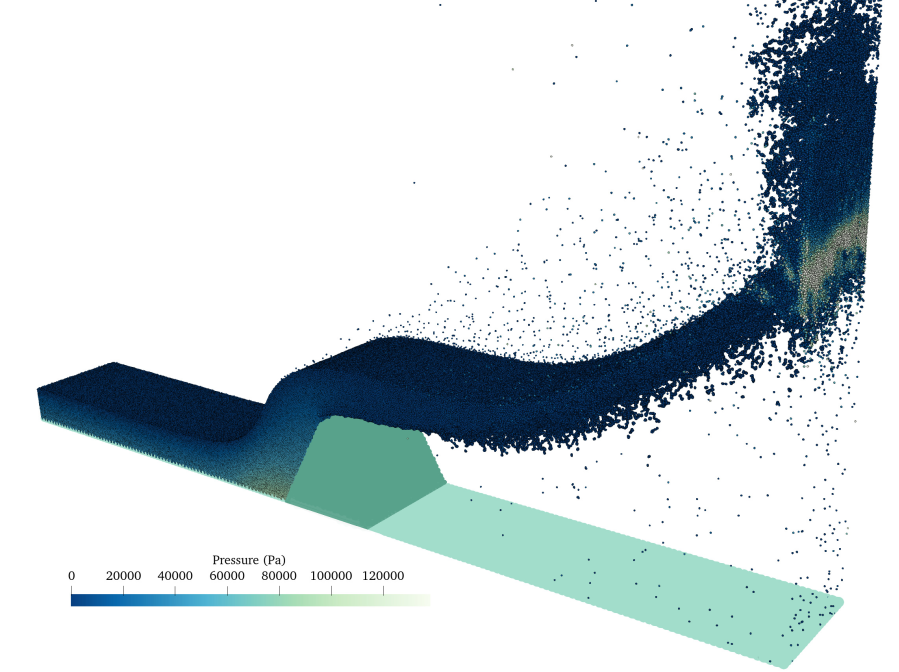}
	\includegraphics[trim = 0mm 0mm 0mm 0mm, clip,width=0.485\textwidth]{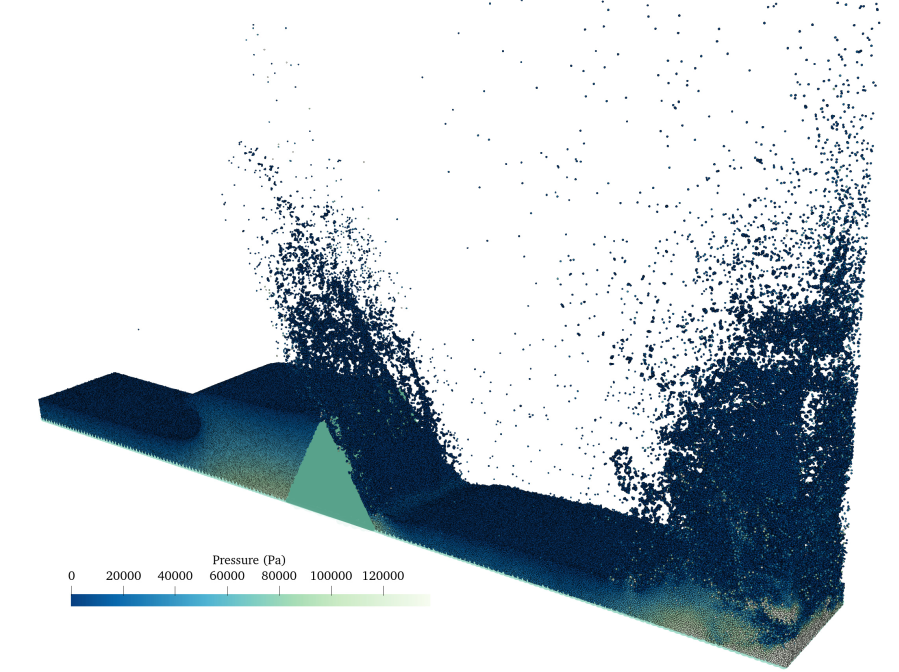}
	\caption{3-D dam break with a wedge: 
		snapshots of the flow evolution together with pressure distribution within the domain.
		The snapshots illustrate the time frames of $t=0,~4,~6$ and $8$ s and $dx = 0.138~$m.}
	\label{figs:dambreakWedgeSnapshots_pressure}
\end{figure}

\begin{figure}
	\centering	
	\includegraphics[trim = 0mm 0mm 0mm 0mm, clip,width=0.485\textwidth]{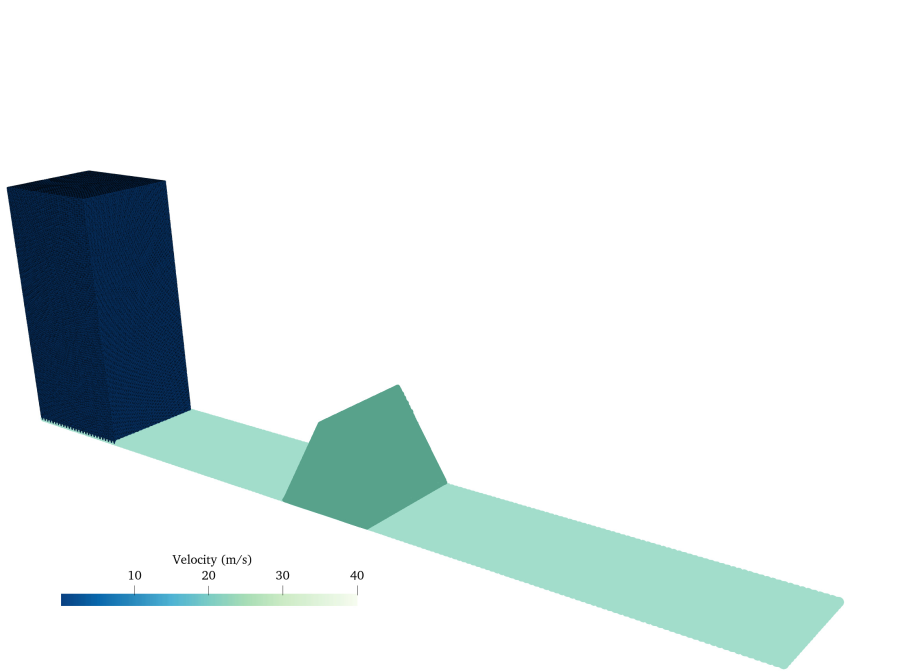}
	\includegraphics[trim = 0mm 0mm 0mm 0mm, clip,width=0.485\textwidth]{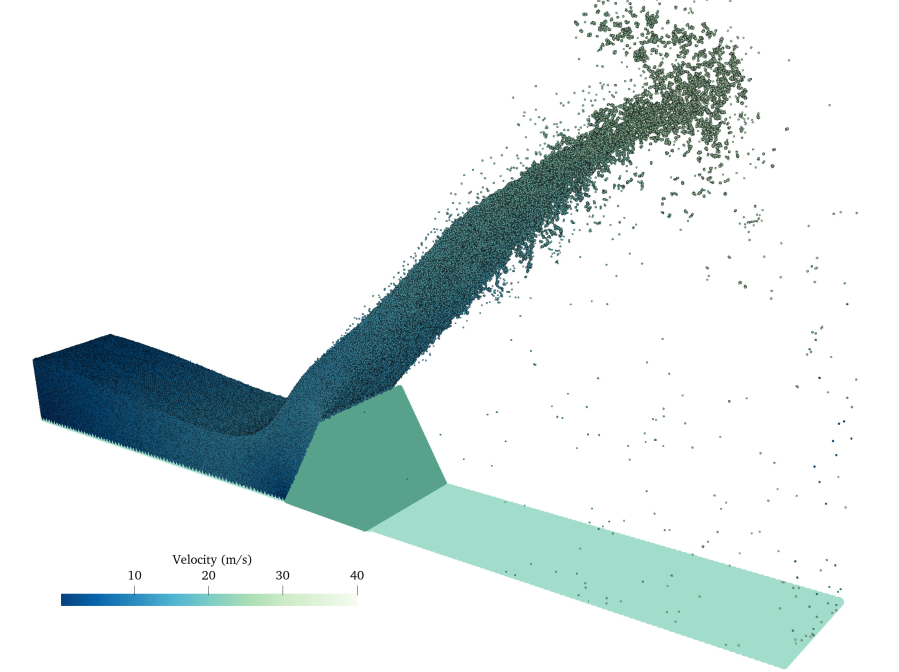}\\
	\includegraphics[trim = 0mm 0mm 0mm 0mm, clip,width=0.485\textwidth]{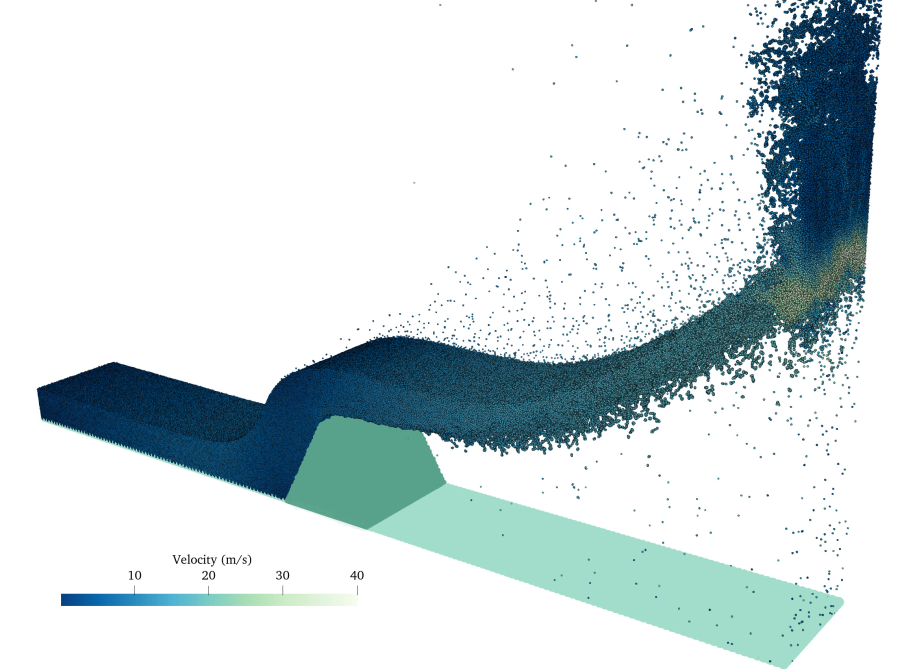}
	\includegraphics[trim = 0mm 0mm 0mm 0mm, clip,width=0.485\textwidth]{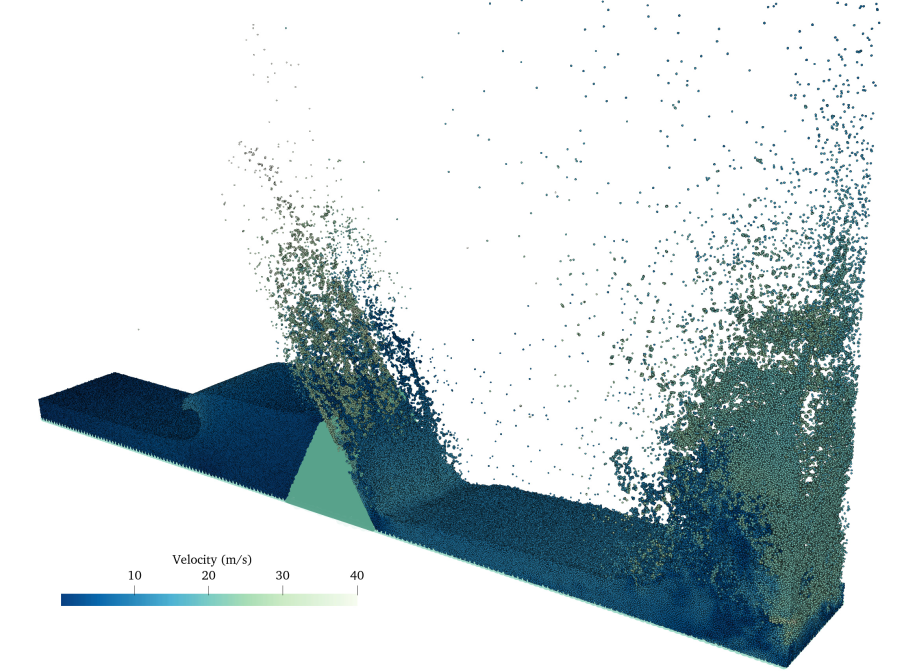}
	\caption{3-D dam break with a wedge: 
		snapshots of the flow evolution together with velocity distribution within the domain.
		The snapshots illustrate the time frames of $t=0,~4,~6$ and $8$ s and $dx = 0.138~$m.}
	\label{figs:dambreakWedgeSnapshots_velocity}
\end{figure}

The evolution of the breaking dam flow and its interactions with the wedge are 
illustrated in several snapshots in figures \ref{figs:dambreakWedgeSnapshots_pressure} 
and \ref{figs:dambreakWedgeSnapshots_velocity}, with the contours of pressure and velocity fields being shown, respectively.
All physical features of the flow are very well captured and the distribution of both pressure and velocity fields are
quite smooth. The interaction of 
water with the wedge at $t=4~$s and also the water jet formed after the 
wedge are effectively simulated.  
The above observations demonstrate
the effectiveness and accuracy of the proposed wall boundary treatment.
As it can be also observed, the large number of particles has resulted in simulations 
with more details of the flow structure.

\begin{figure}
	\centering
	\subfigure []{\includegraphics[width=0.46\linewidth]{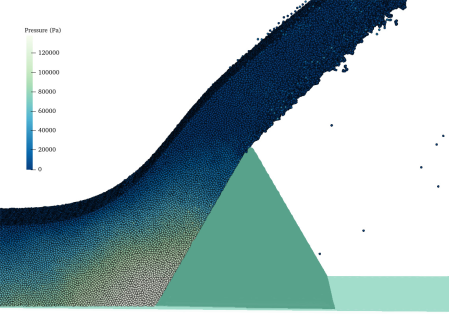}} 
	\hspace{0.3cm}
	\subfigure []{\includegraphics[width=0.46\linewidth]{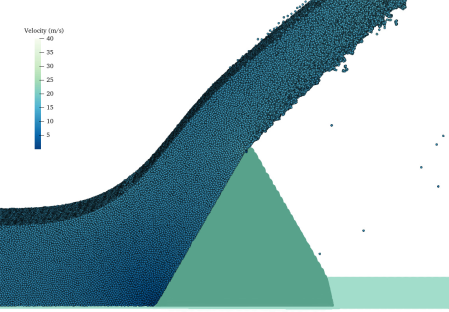}}
	\caption{3-D dam break with a wedge: zoom-in views of the (a) pressure field and (b) velocity
		field at $t=4$ s where the water flow interacts with the wedge and a high-speed 
		water jet is developed.}
	\label{figs:damWedgeZoomInSnapshots}
\end{figure}

The interaction of the high-speed water flow with the wedge is illustrated in more 
details in figure \ref{figs:damWedgeZoomInSnapshots}. It can be seen that
the free-surface, particle splashing and the interaction with wedge are
accurately simulated. The pressure and velocity fields are also quite smoothly
captured. It is also worth mentioning that the effect of the sharp upper tip of the wedge on the flow pattern 
is very well noted, where it deviates the water particles from their relatively
straight trajectory led by the water jet. 

\section{Concluding remarks}\label{sec:conclusions}
In this study, we have presented a robust and efficient wall boundary condition for the SPH method which is accurately
generalised for arbitrary 3-D complex geometries, multi-phase flows, and highly dynamic
problems. As it has been concluded by \citet{valizadeh2015study}, the 
wall boundary condition treatment presented by \citet{ADAMI2012wall} in combination with a density diffusion
formulation achieves the most satisfactory results among the widely used
methodologies in the field of SPH. The presented method in this 
paper, generalises the method of \citet{ADAMI2012wall} for
complex 3-D problems, as well as multi-phase flows,
while improves its accuracy, robustness and efficiency. 

The proposed methodology is also implemented using CUDA to be
executed on GPUs. Various specifications of the method are discussed and profiling results of the GPU-accelerated framework
are summarized. 
The analysis highlights that the 
GPU kernel functions corresponding to the proposed method impose
negligible computational overhead and memory caching, which
identifies that the proposed methodology is ideally suited to the heterogeneous architecture
of GPUs.

The proposed method is validated against existing experimental data, previous numerical 
studies and analytical solutions for both single- and multi-phase problems. 
It is shown that accurate solutions are obtained and potential non-physical particle
penetration into the walls are prevented. Furthermore, it is evidently exhibited
that in predicting violent impact events, the proposed method outperforms the previous methodologies 
of \citet{rezavand2020weakly} and \citet{zhang2020dual},
which have employed the wall boundary treatment proposed by \citet{ADAMI2012wall}. 
Concerning arbitrary configurations with sharp geometries, the proposed method
performs also very well and obtains accurate simulation, reproducing 
substantial features of the flows and fluid-wall interactions. 

As the presented wall boundary condition exhibits unprecedented accuracy and is also efficiently executable
on GPUs with no additionally imposed computational overhead, the method is a 
reliable solution for the long-lasting challenge of the wall boundary condition in the SPH method
for a broad range of natural and industrial applications.

\section*{Acknowledgements}
The authors gratefully acknowledge the financial support by Deutsche 
Forschungsgemeinschaft (DFG HU1572/10-1, DFG HU1527/12-1) for the present work.

\bibliographystyle{jfm}
\bibliography{bibliography}

\end{document}